%% file: phylogenetic.tex
\documentclass{amsart}

\usepackage[english]{babel}
\usepackage[utf8]{inputenc}

\usepackage{phylogenetic}

\def\mytitle{Computing the posterior expectation of phylogenetic trees}
\def\myshorttitle{Posterior expectation of phylogenetic trees}
\def\mykeywords{Bayesian statistics; BHV tree space; \Frechet mean; geometric median; phylogenetic trees; posterior expectation.}
\def\myaffiliation{Max Planck Institute for Mathematics in the Sciences, Inselstr.~22, 04103 Leipzig, Germany}
\def\mythanks{The research leading to these results has received funding from the European Research Council under the European Union's Seventh Framework Programme (FP7/2007-2013) / ERC grant agreement no 267087.}

\IfAmsart{

\title[]{\mytitle}

\author[P. Benner]{Philipp Benner}
\author[M. Ba\v{c}\'ak]{Miroslav Ba\v{c}\'ak}
\address{\myaffiliation}
\email{philipp.benner@mis.mpg.de}
\email{bacak@mis.mpg.de}

\date{\today}
\thanks{\mythanks}

}

\begin{document}
\IfBiometrika{

\jname{Biometrika}
\copyrightinfo{\Copyright\ 2013 Biometrika Trust\goodbreak {\em Printed in Great Britain}}

\received{July 2013}

\markboth{P. Benner \and M. Ba\v{c}\'ak}{\myshorttitle}

\title{\mytitle}

\author{Philipp Benner \and Miroslav Ba\v{c}\'ak}
\affil{\myaffiliation \email{philipp.benner@mis.mpg.de} \email{bacak@mis.mpg.de}}

\maketitle

}


\begin{abstract}
  %
  %
  Inferring phylogenetic trees from multiple sequence alignments often relies upon Markov chain Monte Carlo (MCMC) methods to generate tree samples from a posterior distribution. To give a rigorous approximation of the posterior expectation, one needs to compute the mean of the tree samples and therefore a sound definition of a mean and algorithms for its computation are required. To the best of our knowledge, no existing method of phylogenetic inference can handle the full set of tree samples, because such trees typically have different topologies.
  %
  %
  We develop a statistical model for the inference of phylogenetic trees based on the tree space due to \citet{billera}. Since it is an Hadamard space, the mean and median are well defined, which we also motivate from a decision theoretic perspective. The actual approximation of the posterior expectation relies on some recent developments in Hadamard spaces \citep{bacak,miller} and the fast computation of geodesics in tree space \citep{owenprovan}, which altogether enable to compute medians and means of trees with different topologies.
  %
  %
  We demonstrate our model on a small sequence alignment.
  %
  %
  The posterior expectations obtained on this data set are a meaningful summary of the posterior distribution and the uncertainty about the tree topology.
\end{abstract}

\IfAmsart{\maketitle}
\IfBiometrika{%
}


\section{Introduction}
%
%
Phylogenetic inference is concerned with the estimation of trees that are meant to reflect the evolutionary history of a set of species. Moreover, such point estimates are instrumental to a variety of other inferential tasks, such as the analysis of ChIP-Seq data for the prediction of regulatory elements \citep{Wasserman2004}. A well motivated statistical model with a sound estimation method is therefore of utmost importance for many applications in computational genetics. A number of methods are already available \citep{Huelsenbeck2001,Guindon2003,Drummond2007,Lartillot2009}. They either search for a maximizer of the posterior or likelihood function, or rely on Markov chain Monte Carlo (MCMC) methods to generate samples from the posterior distribution. In phylogenetic inference, posterior samples are phylogenetic trees and their average is usually not well defined unless all trees have the same topology. Here topology refers to the combinatorial structure of a tree. If one considers only one topology at some point of the estimation task, the computed average inevitably neglects part of the data and is not a good summary of the full posterior distribution. A common approach is to construct a (majority rule) consensus tree \citep{Bryant2003} from MCMC samples, for which some decision theoretic arguments have been proposed \citep{Holder2008, huggins} based on the work of \citet{Barthelemy1986}. However, this method is disputed \citep{Wheeler2008} and a more rigorous approach is still lacking.

A first step towards solving this issue was made by \citet{billera} who introduced a space of trees, now called the \emph{BHV tree space,} or simply \emph{tree space,} where a point in this space not only identifies the tree topology, but also the edge lengths. We construct a posterior distribution on this space and show how the expectation and other posterior quantities can be computed. More specifically, since the tree space is an Hadamard space, it admits well-defined notions of a mean and median of probability distributions, which we motivate from decision theoretic grounds. The actual computations of the posterior mean and median rely on approximation algorithms developed by~\citet{bacak,miller}, which in turn require additional tools, mainly the algorithm due to \citet{owenprovan} allowing to compute geodesics between pairs of trees in polynomial time. It is important to emphasize that the construction of the BHV tree space along with the Owen-Provan algorithm provides us with a new way of measuring distances between (phylogenetic) trees, which seem to surpass the conventional metrics (e.g. the NNI distance or the Robinson-Foulds distance) at both mathematical and computational aspects.

%
%

In the present paper, we will give a full description of phylogenetic inference. After a short decision theoretic motivation (Section~\ref{sec:decision}) we will outline the BHV tree space in Section~\ref{sec:ts}, which our model is defined on. To construct a distribution on this space (Section~\ref{sec:sm}), as an intermediate step we first fix a tree topology and thereby restrict the discussion to one orthant of tree space, say the $i$-th orthant. We construct a posterior distribution $\mu_i$ on this orthant, which defines the probability of phylogenetic trees of this topology given a multiple sequence alignment. The posterior distribution $\mu$ on the full tree space is then obtained by combining the single components $\mu_i$, i.e. $\mu \as \sum_i w_i\mu_i$. The main obstacle of this model is the evaluation of the weights $w_i$ since they depend on the partition function of the individual distributions $\mu_i$, which involves computing an intractable integral. We therefore approximate $\mu$ with a finite combination of Dirac measures $\pi$ representing~$K$ samples from the posterior distribution~$\mu.$ To obtain samples from $\mu$ a Markov chain Monte Carlo (MCMC) method is used, which we will describe in Section~\ref{sec:mcmc}.

%
%

Even though our target reader is primarily a practitioner in computational genetics whom we provide with a detailed recipe for a rigorous approximation of posterior distributions in phylogenetic inference, we would like to point out that the presented methods stem from a fascinating mix of pure mathematics including non-Euclidean geometry, convex analysis, optimization, probability theory and combinatorics, which has recently attracted a great deal of interest among mathematicians and keeps offering challenging mathematical problems.


\section{Decision theoretic motivation}
\label{sec:decision}
The goal of any inferential task is to obtain predictions based on a well motivated statistical model and a set of observations. In genetics, such predictions often rely on a phylogenetic tree, which has to be estimated first. Assuming that we already have a posterior distribution $\mu$ on phylogenetic trees, we need to decide on how to obtain a point estimate. Such a tree should be a good summary of the observed data. For this, it is necessary to define a loss function $\loss(s, t)$ \citep{Berger2004,Schervish1995,Robert2001} that quantifies the error of selecting a tree $s$ if $t$ would be a better choice. To illustrate this, assume for the moment that $\Theta$ is a real valued random variable with posterior distribution $\mu_{\Theta\given \Xb}$ conditional on some observations $\{\Xb = \xb\}$. On the real line a common choice is the squared-error loss $\loss(\theta', \theta) = |\theta' - \theta|^2$. As best estimate we would take the minimum expected loss
\begin{equation*}
  \hat{\theta}
  =
  \argmin_{\theta' \in \mathbb{R}}
  \int_{\mathbb{R}}
  \loss(\theta', \theta)
  \di\mu_{\Theta\given \Xb}(\theta \given \xb)
  \ ,
\end{equation*}
and by differentiating with respect to $\theta'$ we immediately find that the estimate $\hat{\theta}$ is the first moment of $\mu_{\Theta\given \Xb}$, i.e.
\begin{equation*}
  \hat{\theta}
  =
  \E(\Theta \given \Xb = \xb)
  =
  \int_{\mathbb{R}}
  \theta
  \di\mu_{\Theta\given \Xb}(\theta \given \xb)
  \ .
\end{equation*}
Similarly we can choose $\loss(\theta', \theta) = |\theta' - \theta|$ for which we obtain the median, whereas a zero-one loss results in a maximum a posteriori (MAP) estimate.

Except for the zero-one loss, such functions are not well defined since trees might be of different topology, but we may take a much more direct approach. As we will outline later, the posterior distribution $\mu$ of our model is defined on tree space $\ts_n$. In this space, all trees have $n+1$ leaves. By definition, $\ts_n$ is a geodesic metric space, which means that we have a metric $d(s, t)$ that defines the distance between $s$ and $t$ and we also have a geodesic path from $s$ to $t,$ whose length is equal to $d(s, t),$ see Section~\ref{sec:ts}. Actually computing the distance involves finding a geodesic path that connects the two trees, which we will discuss later. A possible choice for the loss function is for instance $\loss(s, t) \as d(s, t)^2$. We then obtain the estimate
\begin{equation*}
  \hat{t}
  =
  \argmin\limits_{s\in\ts_n}
  \int_{\ts_n}
  d(s, t)^2
  \di\mu(t)
  \ ,
\end{equation*}
which is also called the barycenter $b(\mu)$ of the distribution $\mu,$ or the \Frechet mean. In Euclidean spaces it coincides with the posterior expectation, which is why we define
\begin{equation*}
  \E(T)
  \as
  b(\mu)
  \ ,
\end{equation*}
where $T$ is a random variable on tree space with distribution $\mu$. For more details on probability theory in Hadamard spaces, see~\citet{sturm,sturm-conm}. We will also use
\begin{equation*}
  \Var(T)
  \as
  \min\limits_{s\in\ts_n}
  \int_{\ts_n}
  d(s, t)^2
  \di\mu(t)
\end{equation*}
as a notion of variance. Similarly, we can choose $\loss(s, t) \as d(s, t)$ and thereby obtain
\begin{equation*}
  \hat{t}
  =
  \argmin\limits_{s\in\ts_n}
  \int_{\ts_n}
  d(s, t)
  \di\mu(t)
\end{equation*}
as estimate, which is the geometric median. The advantage of the geometric median is that it is less sensitive to long tails of the distribution, but it may not have a unique minimizer. Since the distance function on an Hadamard space is convex, computing a point estimate of $\mu$ reduces to finding a minimizer of a convex function. In tree space, we cannot simply differentiate the loss function and follow the gradient to find a minimizer. Appropriate algorithms for computing the mean and median are referred to in Section~\ref{sec:mcmc}.

Of course, a valid question is whether a single point estimate $\hat{t}$ is a good summary of the full posterior distribution $\mu$. In real applications part of the data will favour one tree topology, while another part clearly supports a different topology. Such seemingly contradictory data sets are very frequent in biological applications and lead to posterior distributions whose mass sits on many topologies. In this case a weighted mixture of several trees, i.e. a model average, might be a better summary of the posterior distribution. When computational time is a limiting factor, it might be too costly to use a model average. However, we will demonstrate in Section~\ref{sec:results} that also single point estimates in tree space may allow an intuitive interpretation of multimodal posteriors. We would like to mention an alternative approach due to~\citet{nye} which instead of a point estimate uses \emph{principal component analysis} in tree space.


\section{Phylogenetic trees and tree space}
\label{sec:ts}
\begin{figure}[ht]
\centering
\includegraphics[width=50mm]{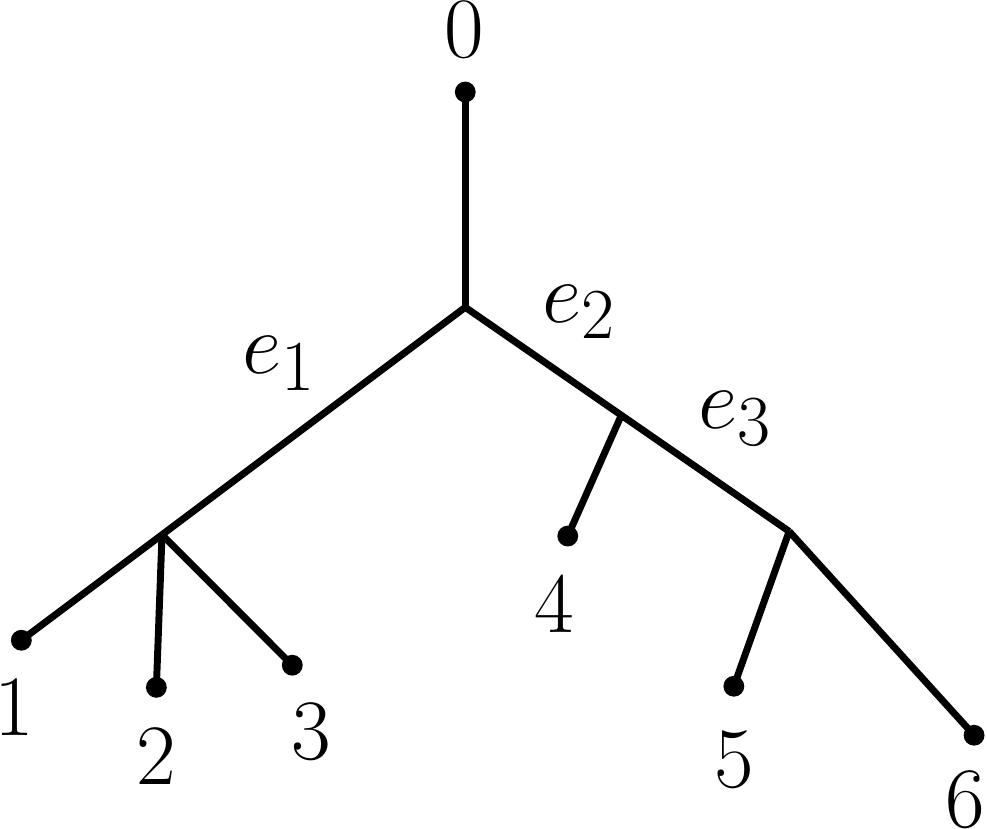}
\caption{An example of a $6$-tree with three inner edges.}
\label{fig:tree}
\end{figure}

We will now describe the construction of tree space due to L.~Billera, S.~Holmes, and K.~Vogtmann. For the details, the interested reader is referred to the original paper~\citet{billera}. We first need to make precise what we mean by a (phylogenetic) tree. Given $n\in\nat$ with $n\geq3,$ a \emph{metric $n$-tree} is a combinatorial tree (connected graph with no circuit) with $n+1$ terminal vertices called \emph{leaves} that are labeled $0,1,\dots,n.$ In phylogenetics, the labels represent the species in question. The vertex connected with leaf $0$ is called the \emph{root}, since it represents a common ancestor of all species in the tree, but it will have no distinguished role in the construction of tree space. (As a matter of fact, such trees can be considered as unrooted.) Some authors however use the term root for the leaf vertex $0$ itself. Vertices other than leaves have no labels since we view them just as ``branching points''. The edges which are adjacent to leaves are called \emph{leaf edges,} and the remaining edges are called \emph{inner.} We see an example of a $6$-tree with three inner edges $e_1,e_2,$ and $e_3$ in Fig.~\ref{fig:tree}.
All edges, both leaf and inner, have positive lengths. We will refer to a metric $n$-tree simply as a \emph{tree.} The number $n$ will be fixed and clear from the context. Later, when we consider a \emph{set} of trees instead of an individual tree, it will be important that they all have the same number of leaves. For the inference of phylogenetic trees, the number of leaves is determined by the number of nucleotide sequences in the data set.

Each inner edge of a tree determines a unique partition of the set of leaves $L$ into two disjoint and nonempty subsets $L_1\cup L_2=L$ called a \emph{split,} which we denote $L_1| L_2.$ A split is defined as the partition of leaves that arises if we removed the inner edge under consideration. For instance, the inner edges $e_1,e_2,$ and $e_3$ of the tree in Fig.~\ref{fig:tree} have splits $(0,4,5,6|1,2,3), \;(0,1,2,3|4,5,6),$ and $(0,1,2,3,4|5,6),$ respectively. On the other hand, given a set of leaves and splits subject to certain conditions, we can uniquely construct a tree. Namely, we require that any two splits $L_1| L_2$ and $L_1'| L_2'$ are \emph{compatible,} that is, one of the sets
\begin{equation*} 
 L_1\cap L_2',\quad L_1'\cap L_2,\quad L_1\cap L_1',\quad L_2\cap L_2' 
\end{equation*}
must be empty. We say that a set of inner edges $I$ is \emph{compatible} if for any two edges $e,e'\in I,$ the corresponding splits are compatible. For further details, see~\citet{dress,semple}.

We will now proceed to construct a space of trees, denoted $\ts_n,$ that is, a space whose elements will be all metric $n$-trees. First, it is useful to realize that one can treat leaf edges and inner edges separately. Since the former can be represented in Euclidean space of dimension $n+1,$ the whole space $\ts_n$ is a product of a Euclidean space and a space that represents the inner edges. We may hence for simplicity ignore the leaf edges in the following construction.

Fix now a metric $n$-tree $t$ with $r$ inner edges of lengths $l_1,\dots,l_r,$ where $1\leq r\leq n-2.$ Clearly $\left(l_1,\dots,l_r\right)$ lies in the open orthant $(0,\infty)^r,$  and conversely, any point of $(0,\infty)^r$ can be mapped to an $n$-tree of the same combinatorial structure as $t.$ Note that a tree $S$ is said to have the same combinatorial structure as $t$ if it has the same number of inner edges as $t$ and all its inner edges have the same splits as the inner edges of $t.$ In other words, the trees $s$ and $t$ differ only by inner edge lengths.

To any point of the boundary $\partial (0,\infty)^r$ we associate a metric $n$-tree obtained from $t$ by shrinking some inner edges to zero length. Hence, each point from the closed orthant $[0,\infty)^r$ corresponds to a metric $n$-tree of the same combinatorial structure as $t.$ 

\begin{figure}[ht]
  \centering
  \includegraphics[width=70mm]{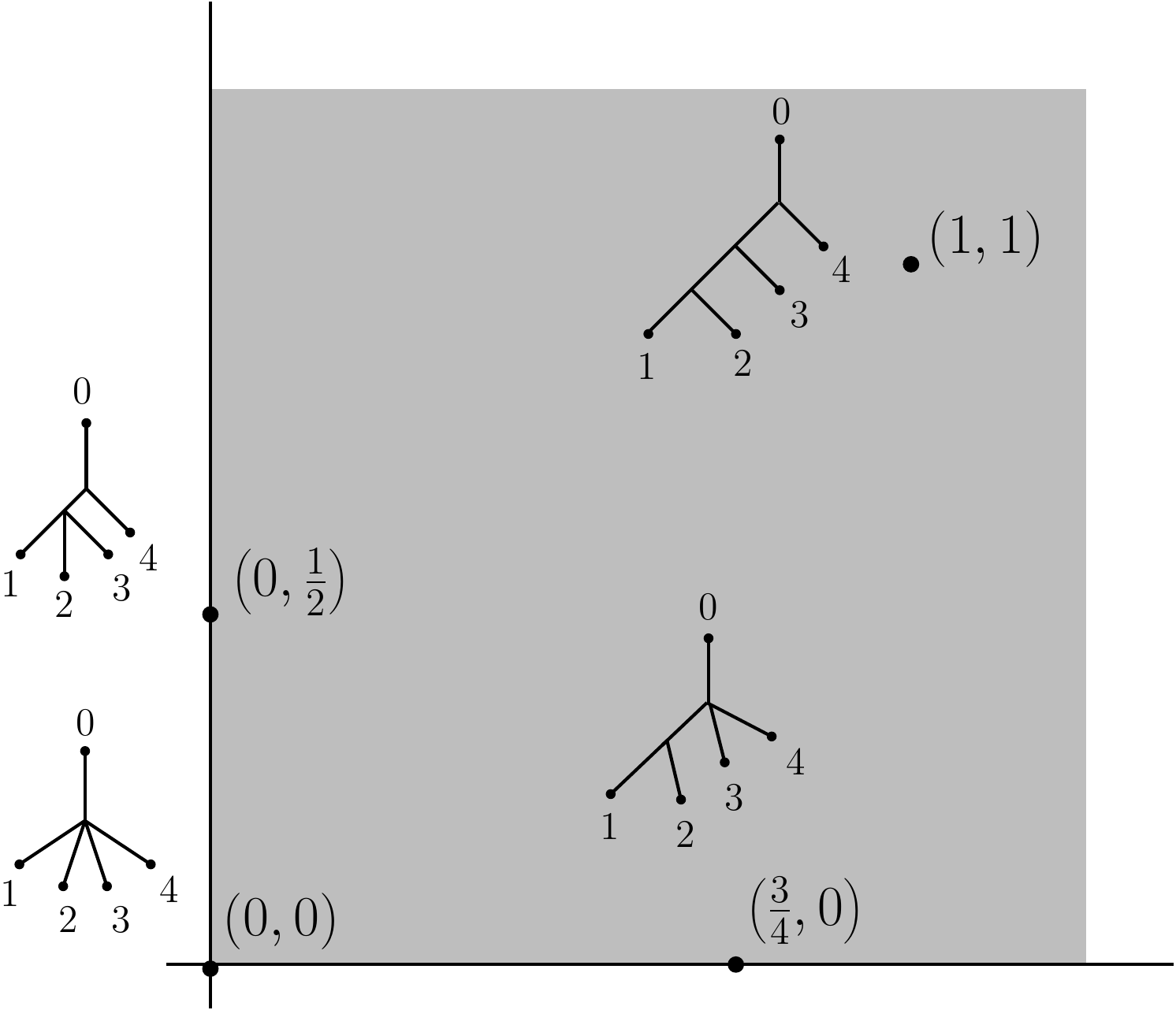}
  \caption{$4$-trees of a given combinatorial structure.}
  \label{fig:orthant}
\end{figure}

Binary $n$-trees have the maximal possible number of inner edges, namely $n-2,$ which is of course equal to the dimension of the corresponding orthant. An orthant of an $n$-tree that is not binary appears as a face of the orthants corresponding to (at least three) binary trees. In Fig.~\ref{fig:orthant}, we see a copy of $[0,\infty)^2$ representing all $4$-trees of a given combinatorial structure, namely, all $4$-trees with two inner edges $e_1$ and $e_2,$ such that the split of $e_1$ is $(1,2|0,3,4),$ and the split of $e_2$ is $(1,2,3|0,4).$ If the length of $e_1$ is zero, then the tree lies on the vertical boundary ray. If the length of $e_2$ is zero, then the tree lies on the horizontal boundary ray.
In summary, any orthant $\co=[0,\infty)^r,$ where $1\leq r\leq n-2,$ corresponds to a compatible set of inner edges, and conversely, any compatible set of inner edges $I=\left(e_1,\dots,e_r\right)$ corresponds to a unique orthant $\co(I),$ which is a copy of $[0,\infty)^r.$

The \emph{tree space} $\ts_n$ consists of $(2n-3)!!\as(2n-3)(2n-5)\cdot\dots\cdot5\cdot3$ copies of the orthant $[0,\infty)^{n-2}$ glued together along lower-dimensional faces, which correspond to non-binary trees, that is, compatible sets of inner edges of cardinality $<n-2.$

\begin{figure}[ht]
  \centering
  \subfigure[]{
    \def\svgwidth{0.25\textwidth}
    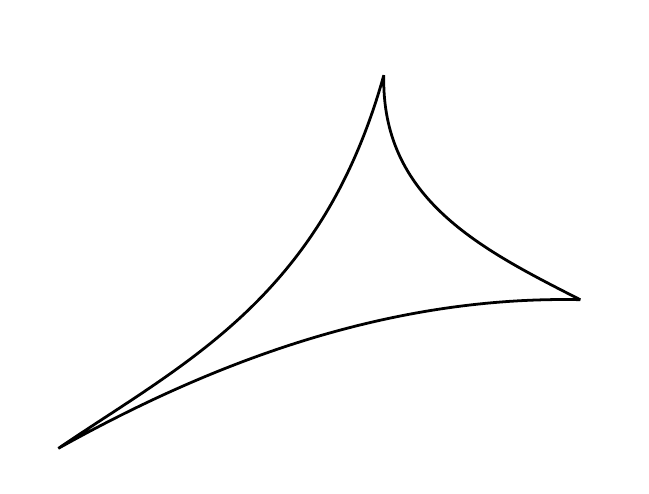}
  \hspace{1cm}
  \subfigure[]{
    \def\svgwidth{0.25\textwidth}
    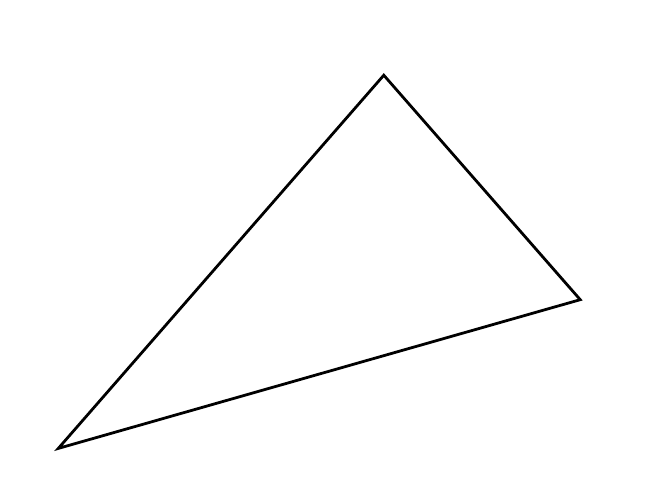}
  \caption{(a) Triangle in a space of nonpositive curvature. (b) Comparison triangle in Euclidean space.}
  \label{fig:triangle}
\end{figure}

We equip the tree space $\ts_n$ with the induced length metric. Then it becomes a geodesic metric space, that is, given a pair of trees, we have a well-defined distance between them and moreover they are connected by a geodesic path. One can easily observe that each geodesic consists of finitely many Euclidean line segments. An algorithm for the computation of distances and geodesics is due to \citet{owenprovan}. The following important theorem from~\citet{billera} states that the tree space has nonpositive curvature.
\begin{thm}
The space $\ts_n$ is an Hadamard space.
\end{thm}
An Hadamard space is a geodesic metric space, which is complete and has nonpositive curvature. Intuitively, in such spaces triangles appear ``slimmer'' than in Euclidean space, see Fig.~\ref{fig:triangle}.

\begin{figure}[ht]
  \centering
  \includegraphics[width=70mm]{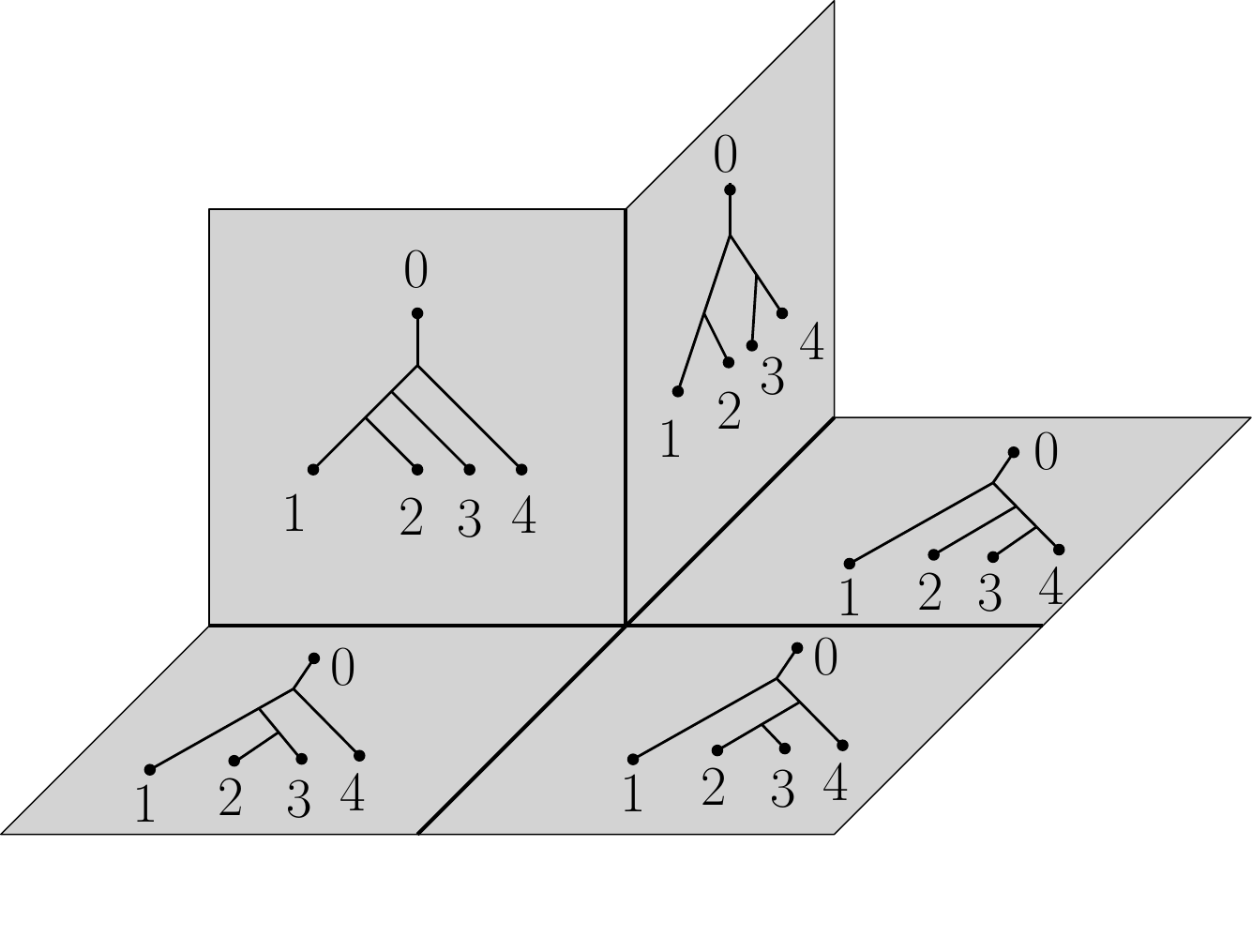}
  \caption{Five out of $15$ orthants of $\ts_4.$}
  \label{fig:t4}
\end{figure}

It is impossible to isometrically embed the tree space into the Euclidean space and therefore also difficult to visualize. A piece of the tree space $\ts_4$ is shown in Fig.~\ref{fig:t4}. The geometrically oriented reader may notice that triangles in this space are deformed and squeezed inwards, that is, they are ``slim'' as explained above.

Since this space is not a linear space, addition of two elements of tree space is also not defined. However, convex combinations of a given pair of points are meaningful. Indeed, let $s,t\in\ts_n$ and $\lambda\in[0,1],$ then we define a formal convex combination
\begin{equation*}
 t_\lambda \as (1-\lambda)s+\lambda t,
\end{equation*}
which represents a unique tree $t_\lambda\in\ts_n$ lying on the geodesic from $s$ to $t$ satisfying $d\l(s,t_\lambda\r)=\lambda d\l(s,t\r).$ Convex combinations are important in the algorithms for computations of medians and means.


\section{Statistical model}
\label{sec:sm}
The posterior distribution is a conditional probability measure that depends on a multiple sequence alignment. Such an alignment is represented as a matrix, where each row is a sequence of nucleotides from one species. Within each column (site), observed nucleotides are assumed to have evolved according to a phylogenetic tree. The same tree is assumed for the whole alignment. A most intuitive way to describe this process is to look at it from a generative model perspective. We assume that there existed a common ancestor of all species that we are considering and the nucleotides that we observe are generated from the sequence of the common ancestor. Whenever a mutation occurs between an ancestor and its descendant, a new nucleotide is generated from the stationary distribution of the process, at which point it might happen that the same nucleotide is generated again. The stationary distribution therefore plays a crucial role. It is specific to each column of the alignment and reflects the external selective pressure that acts on each site. This type of model was for instance also used by \citet{Siddharthan2005}. Other more commonly used methods assume the same stationary distribution among all sites. In some methods it is possible to group sites into distinct classes that share a stationary distribution \citep{Lartillot2004}. To explicate the differences to other methods and how prior parameters should be interpreted we fully outline the model in the following. However, any other model might be used and it is not important to the later discussions of this paper.

For a more formal description it is sufficient for the first part to develop the statistical model on the set of observations within a single column of the alignment. Let $n+1$ be the number of species for which we have sequences in the alignment. We introduce the random variables $\Xb = \{ X_0, \dots, X_n \}$, where $X_i$ takes values in an alphabet $\al$ and represents the nucleotide of the $i$-th sequence. The alphabet contains a character for each nucleotide and one to represent gaps in the alignment. By including a symbol for gaps in the alphabet we explicitly state that no nucleotide is present at positions filled with a gap. If however gaps are modeled as missing data, the meaning of gaps is different, i.e. a gap indicates that any of the nucleotides is present but which one is unknown. One of the sequences in the alignment is used for the outgroup, for which we use the random variable $X_0$ associated with the leaf which is attached to the root of the tree. A phylogenetic tree equipped with an evolutionary model is used to relate sequences of different species. We first consider a particular phylogenetic $n$-tree $T$ with $r = n-2$ inner edges. The leaves of the tree $\{v_0, \dots, v_n\}$ are associated with the $n+1$ random variables $\Xb$. Inner vertices are labeled from $n+1$ to $n+r+1$ and we associate with each inner vertex $v_k$ a random variable $X_k$, where $k \in \{n+1, \dots, n+r+1\}$. To discuss the evolutionary model, assume that $v_i$ and $v_j$ are leaves or inner vertices that are connected to the $k$-th inner vertex $v_k$. We need to define the probability of an event $\{X_i = x_i, X_j = x_j\}$ knowing that $\{X_k = x_k\}$. First we assume that
\begin{equation*}
  X_i \independent X_j \given X_k
\end{equation*}
and therefore the conditional probability of $\{X_i = x_i, X_j = x_j\}$ given $\{X_k = x_k\}$ factorizes. Hence, it is sufficient to specify the probability of $\{X_i = x_i \}$ given $\{X_k = x_k\}$. We use the model by \citet{Felsenstein1981}, which defines a continuous-time finite Markov chain. It is given by
\begin{equation*}
  \p_{X_i \given X_k}(x_i \given x_k) \as
  \begin{cases}
    \mut{i} \stationary(x_i) + \nomut{i} & \text{if} \ x_i = x_k \ ,\\
    \mut{i} \stationary(x_i)             & \text{if} \ x_i \neq x_k \ ,\\
  \end{cases}
\end{equation*}
where $\mut{i}$ is the probability of a mutation and $\nomut{i} = 1-\mut{i}$. This substitution model is also called F81+Gaps, see for instance \citet{Mcguire2001}. The distribution \stationary is the stationary probability distribution of the process, which is common to the full tree. In this model, the case where $x_i \neq x_k$ is simple, we have a mutation and generate the nucleotide $x_i$ with probability $\stationary(x_i)$. More interestingly, if $x_i$ and $x_k$ are the same nucleotides, there is either no mutation and no nucleotide has to be generated or there is a mutation and the same nucleotide is generated again. As we will outline later, the entropy of \stationary defines the level of conservation of a site. More complex substition models exist that for instance account for differences in transitions and transversions \citep[e.g.][]{Hasegawa1985}. However, we stick to the simpler F81+Gaps model for mathematical simplicity and in order not to overparameterize the statistical model, since we already use a site-specific stationary distribution. The probability of a mutation $\mut{i}$ depends on the distance $l$ between species $v_i$ and its ancestor $v_k$, but also on the evolutionary rate $\omega$. In Felsenstein's evolutionary model, we set
\begin{equation*}
  \mut{i} \as 1-e^{-\omega l} \ ,
\end{equation*}
where $\omega$ is the same for all edges. The rate parameter is often assumed to be column specific \citep[e.g.][]{Yang1993} and used to control the level of conservation. As pointed out later, we control the level of conservation with the entropy of the stationary distribution and therefore set $\omega = 1$ for all columns in the alignment. The mutation model is time-reversible, which means that inference is restricted to unrooted trees. This property allows us to define our statistical model on the BHV tree space. As discussed in Section~\ref{sec:ts}, our phylogenetic trees have at least three vertices attached to the root, which essentially makes the tree unrooted. The leaf associated with $X_0$ can be seen to represent an outgroup. The position of the root is purely instrumental and has no importance for the computation of the likelihood \citep{Isaev2006}. Since no observations are available for the inner vertices, it is necessary to marginalize over all corresponding random variables, such that for instance
\begin{equation*}
  \p_{X_i, X_j}(x_i, x_j)
  =
  \sum_{x_k \in \al}
  \stationary(x_k)
  \p_{X_i, X_j \given X_k}(x_i, x_j \given x_k)
  \ .
\end{equation*}
In this fashion we obtain the full likelihood of $\{ \Xb = \xb \}$.

In this model we have two sets of unobserved parameters, namely the lengths of edges and the stationary probability distribution. We proceed by discussing the stationary distribution first, which is specific to each column of the alignment. The distribution will be integrated out in the full model, since we are only interested in the inference of phylogenetic trees. We introduce a random variable $\Theta$ that represents the stationary distribution and obtain the conditional probability
\begin{equation*}
  \p_{X \given \Theta} (x \given \varthetab)
  =
  \vartheta_x
  =
  \stationary(x)
  \ ,
\end{equation*}
of generating nucleotide $x \in \al$. We assume that $\Theta$ is a priori Dirichlet distributed with pseudocounts $\alphab = (\alpha_x)_{x \in \al}$. The probability of observing $\{ \Xb = \xb \}$ becomes
\begin{equation*}
  \p_{\Xb}(\xb)
  =
  \int_{\Delta}
  \p_{\Xb \given \Theta}(\xb \given \varthetab)
  f_{\Theta}(\varthetab)\di\varthetab
  \ ,
\end{equation*}
where $f_\Theta$ is the density function of the Dirichlet distribution. The integral is defined on the $\l(|\al|-1\r)$-dimensional probability simplex $\Delta$ and can be solved analytically by first expanding the polynomial of the distribution $\p_{\Xb \given \Theta}$.

It is important to select an appropriate set of parameters $\alphab$ for the Dirichlet distribution, as they control the expected entropy of distributions $\varthetab$ drawn from it. Phylogenetic trees are commonly learned on multiple sequence alignments of genes. Such genomic regions are highly conserved, which means that selective pressure causes nucleotides in a column of the alignment to be the same with high probability. To reflect this knowledge in our prior assumption, it is important that the expected entropy is low, i.e. that only the probability of one or two nucleotides is high. This can be achieved by choosing $\alpha_x < 1$, which puts mass on the boundaries of the probability simplex. The choice of $\alphab$ has a strong influence on inferred edge lengths. If we increase $\alphab,$ we observe that inferred branch lengths shorten to compensate for the increase in entropy of the stationary distribution. The choice of pseudocounts $\alphab$ therefore reflects our \emph{a priori} assumption of how conserved we expect a genomic region to be. It is well known that within codons a heterogeneous selective pressure exists \citep{Li1985,Yang1996}, which can be modeled by introducing specific pseudocounts.

The next step is to formulate a prior distribution on the edge lengths given a fixed topology. By this we obtain the posterior $\mu_i$ for a single orthant $\orthant_i$. The same phylogenetic tree is assumed for all columns in the alignment. In fact, columns in the alignment are conditionally independent given a fixed phylogenetic tree. Since we now want to let the tree vary within one orthant of tree space, it is necessary to consider the full alignment. Let $\Xb^{(\nu)}$, where $\nu = 1, \dots, N$, denote the random variables for the $\nu$-th column of the alignment. We also use the shorthand notation $\ol{\Xb} = (\Xb^{(1)}, \dots, \Xb^{(N)})$ for the full alignment. Let $\Lb = (L_k)$ denote the random variables for the edge lengths of a tree $T$ in orthant $\orthant_i$. Each $L_k$ is a priori gamma distributed with shape parameter $b$ and scale parameter $\lambda$. The likelihood of the full alignment is given by
\begin{equation*}
  \p_{\ol{\Xb} \given \Lb, \orthant_i}
  (\ol{\xb} \given \lb)
  =
  \prod_{\nu=1}^N
  \p_{\Xb^{(\nu)} \given \Lb, \orthant_i}
  (\xb^{(\nu)} \given \lb)
  \ ,
\end{equation*}
where the stationary distribution is integrated out, and we obtain the posterior distribution $\mu_i$ restricted to orthant $\orthant_i$ with density function 
\begin{equation*}
  f_{\Lb \given \ol{\Xb}, \orthant_i} (\lb \given \ol{\xb})
  =
  \frac{
    1
  }{
    \p_{\ol{\Xb} \given \orthant_i}(\ol{\xb})
  }
  \p_{\ol{\Xb} \given \Lb, \orthant_i}
  (\ol{\xb} \given \lb)
  f_{\Lb}(\lb)
  \ .
\end{equation*}
The full posterior distribution of $n$-trees is given by
\begin{equation*}
  \mu
  \as
  \sum_{i=1}^{(2n-3)!!} w_i\mu_i
  \ ,
\end{equation*}
where
\begin{equation*}
  w_i
  \as
  \frac{
    \p_{\ol{\Xb} \given \orthant_i}(\ol{\xb})
  }{
    \sum_j \p_{\ol{\Xb} \given \orthant_j}(\ol{\xb})
  }
\end{equation*}
is the weight of the $i$-th component. We will denote the density function of $\mu$ simply as $f$. The weight $w_i$ depends on the normalized partition function of $\mu_i$, which involves computing an intractable integral. Another difficulty is that the number of orthants grows super-exponentially with the number of leaves. It is therefore necessary to approximate $\mu$ with a Dirac mixture of posterior samples, which does not require to compute any partition functions.


\section{Approximation of the posterior distribution}
\label{sec:mcmc}
\begin{figure}[ht]
  \centering
  \subfigure[]{
    \begin{tikzpicture}[>=latex,xscale=0.7,yscale=0.3]
      \input{phylogenetic_mcmc-moves-a.tex}
    \end{tikzpicture}
  }
  \subfigure[]{
    \begin{tikzpicture}[>=latex,xscale=0.7,yscale=0.3]
      \input{phylogenetic_mcmc-moves-b.tex}
    \end{tikzpicture}
  }
  \subfigure[]{
    \begin{tikzpicture}[>=latex,xscale=0.7,yscale=0.3]
      \input{phylogenetic_mcmc-moves-c.tex}
    \end{tikzpicture}
  }
  \caption{Possible MCMC moves to neighboring orthants (nearest-neighbor interchange, NNI). The edge $e$ of tree (a) can be replaced by two other edges in neighboring orthants shown in trees (b) and (c). The leaves labeled from zero to three may also represent more complex subtrees.}
  \label{fig:moves}
\end{figure}
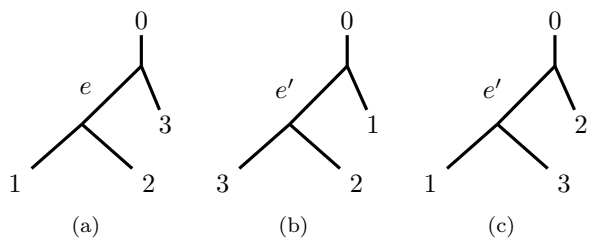

To summarize the posterior $\mu$, we would like to compute a point estimate
\begin{equation*}
  \hat{t}
  =
  \argmin\limits_{s\in\ts_n}
  \int_{\ts_n}
  \loss(s, t)
  \di\mu(t)
  \ ,
\end{equation*}
for an appropriate loss function $\loss$, as discussed in Section~\ref{sec:decision}. Unfortunately, the expected loss is difficult to compute and we therefore rely on an approximation by replacing $\mu$ with the Dirac mixture
\begin{equation*}
  \pi \as \frac{1}{K}\sum_{k=1}^K \delta_{t_k}
\end{equation*}
of $K$ samples from $\mu$. By the ergodic theorem, we have the convergence
\begin{equation*}
  \int_{\ts_n}
  \loss(s, t)
  \di\pi(t)
  =
  \frac{1}{K}\sum_{k=1}^K 
  \loss(s, t_k)
  \to
  \int_{\ts_n}
  \loss(s, t)
  \di\mu(t)
  \ ,
\end{equation*}
almost surely for every $s \in \ts_n$ as $K \rightarrow \infty$ \citep{Robert1999}. A set of posterior samples can be obtained with the Metropolis-Hastings algorithm \citep{Metropolis1953,Hastings1970} without having to evaluate the weights $w_i$ of the single components of $\mu$. The algorithm constructs a Markov chain with $\mu$ as the stationary distribution. Let $t_k$ be a sample from $\mu$ with edge set $\ce$. A new sample $t_{k+1}$ is generated by the Markov chain conditional on the current sample $t_k$. The algorithm uses a~proposal distribution with density function $q(\cdot\given t_k)$, which selects an edge $e \in \ce$ and replaces it by another edge. We thereby obtain a new tree $s$ that we accept as the next sample $t_{k+1}$ with probability
\begin{equation*}
  \rho(t_k, s)
  =
  \min
  \left\{
    1,
    \frac{
      f(s)q(t_k\given s)
    }{
      f(t_k)q(s\given t_k)
    }
  \right\}
  \ ,
\end{equation*}
and otherwise $t_{k+1} = t_k$, where $f$ still denotes the density function of $\mu$. Note that the normalization constant of $\mu$ cancels in the ratio. The proposed tree $s$ lies in the same orthant as $t_k$ with probability $\tau$. In this case, a new edge length is proposed, which is a draw from a normal distribution centered at $|e|$. However, with probability $1-\tau$ the proposed tree lies within one of the neighboring orthants (NNI move), by replacing the edge $e$ by one of two other possible edges of the same length (see Fig.~\ref{fig:moves}). Other MCMC methods also make use of subtree pruning and regrafting (SPR) moves, which allow global jumps in tree space. For the small examples in Section~\ref{sec:results} we believe that NNI moves are sufficient. The transition measure of the Markov chain is given by
\begin{equation*}
  \kappa(x, \di y)
  =
  \rho(x, y)q(y \given x)\di y
  +
  (1-r(x))\delta_{x}(\di y)
\end{equation*}
with $r(x) = \int_{\ts_n} \rho(x, y)q(y \given x) \di y$, which satisfies the detailed balance condition and therefore has $\mu$ as invariant distribution \citep{Robert1999}.

In this paper, we will focus on approximating the mean and median of the posterior distribution. The problem of finding a point estimate therefore reduces to computing the geometric median
\begin{align}
  \label{eq:median}
  \Psi\left(\ol{t}\right) &\as \argmin_{s\in \ts_n} \sum_{k=1}^K d\left(s,t_k\right) \ ,
  \intertext{and the \Frechet mean}
  \label{eq:mean}
  \Xi\left(\ol{t}\right)  &\as \argmin_{s\in\ts_n} \sum_{k=1}^K  d\left(s,t_k\right)^2 \ ,
\end{align}
of a finite set $\ol{t}\as\l(t_1,\dots,t_K\r)$ of trees from $\ts_n.$ Since both the median and mean are defined as minimizers of ``nice'' convex functions on tree space, we get the following. The median $\Psi\l(\ol{t}\r)$ always exists and it is unique unless all the trees $t_1,\dots,t_K$ lie on a geodesic. The existence and uniqueness of $\Xi\l(\ol{t}\r)$ is a consequence of strong convexity of the minimized function. The interested reader is referred to~\citet[Theorem 3.2.1]{jost2} and \citet[Proposition~4.4]{sturm-conm}. The proofs can be also found in~\citet[Theorem 2.4]{bacak}. Note that medians and means are well-defined on arbitrary Hadamard spaces.
 
We will now turn to the question of how to \emph{compute} medians and means of a given set of trees, since the formulas~\eqref{eq:median} and~\eqref{eq:mean} do not provide us with direct algorithms. It turns out that efficient approximation methods from optimization can be extended into Hadamard spaces and applied to median and mean computations. For explicit algorithms, the reader is referred to \citet[Section~4]{bacak}, where a random and a cyclic-order version of an approximation algorithm are presented. Note that we consider \emph{unweighted} medians and means here, which slightly simplifies the formulas in \citet[Section~4]{bacak}. Interestingly, the random-order version of the algorithm for computing the mean can be alternatively justified via the law of large numbers due to \citet{sturm}, as was independently observed by \citet{bacak} and \citet{miller}. For the reader's convenience, the random-order version for unweighted medians and means is outlined in Appendix~\ref{sec:mmalg}.

The approximation algorithms for computing medians and means use (at each step) the algorithm for finding a geodesic in tree space by \citet{owenprovan}. At a crucial stage, the Owen-Provan algorithm computes a maximal flow. Our implementation \citeauthor{tfbayes} does this using an encoding of the max flow problem as an integer program. In contrast, our implementation \citeauthor{trap} uses the push-relabel algorithm due to~\citet{goldberg}.



\section{Results}
\label{sec:results}
To demonstrate our method we used data from the UCSC multiz46way alignment. We selected a subset of the MT-RNR2 gene alignment\footnote{The Hg19 coordinates of the sequence are chrM: 1686-2059.} (16S rRNA, see \citealt{anderson1981}), because the posterior distribution has significant mass on multiple tree topologies. Two example computations are presented in the following. For both, a gamma prior on edge lengths with shape parameter $b = 1$ and scale parameter $\lambda = 0.1$ is used. This choice of parameters reflects our belief that edge lengths can be very small and allows the sampler to easily switch between orthants. On the other hand, a shape parameter of $b > 1$ would cause the posterior to have more distinct modes. The Dirichlet prior on the stationary distribution has pseudocounts $\alpha_x = 0.2$ for all $x \in \al$, which reflects our believe that the data set is a conserved genomic region. The unnormalized log posterior of MCMC samples is shown in Fig.~\ref{fig:ex-posterior} for both examples, which indicates that the Markov chain mixes well. Since we analytically integrate over the stationary distribution of the mutation model, we expect that much less samples are required for a good approximation compared to methods that do otherwise.

\begin{figure}[ht]
  \centering
  \begin{tikzpicture}[>=latex,scale=0.6,x=1pt,y=1pt]
    \input{phylogenetic_example1.mean.tex}
  \end{tikzpicture}
  \caption{Example 1: \Frechet mean of posterior samples with an approximate variance of $0.0248$. Edge lengths are visualized as distances in the horizontal direction only.}
  \label{fig:ex1-mean}
\end{figure}
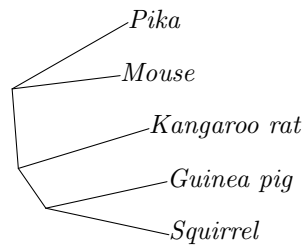

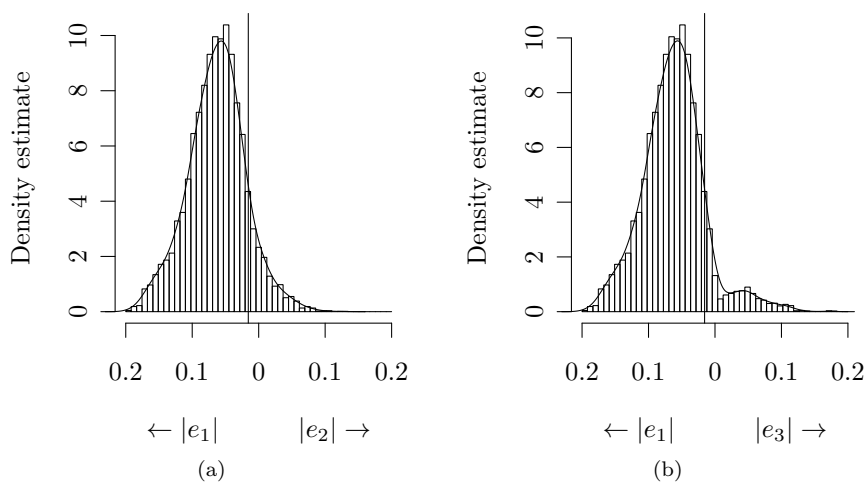
\begin{figure}[ht]
  \centering
  \subfigure[]{
    \label{fig:ex1-hista}
    \begin{tikzpicture}[>=latex,scale=0.9,x=1pt,y=1pt]
      \input{phylogenetic_example1.histogram1.tex}
    \end{tikzpicture}
  }
  \subfigure[]{
    \label{fig:ex1-histb}
    \begin{tikzpicture}[>=latex,scale=0.9,x=1pt,y=1pt]
      \input{phylogenetic_example1.histogram2.tex}
    \end{tikzpicture}
  }
  \caption{Example 1: Marginal posterior density estimate of three edges. The posterior expectation is shown as a vertical line.}
  \label{fig:ex1-hist}
\end{figure}

In the first example we considered only five species, namely Guinea pig, Kangaroo rat, Mouse, Pika, and Squirrel. The approximated posterior expectation was computed on the last 16000 samples and is shown in Fig.~\ref{fig:ex1-mean}. We do not show the median because it is very similar to the mean. Consider the edge
\begin{align*}
  e_1 &: \nsplit{Pika, Mouse}{Kangaroo rat, Squirrel, Guinea pig} \ ,
  \intertext{which connects the subtree of Pika and Mouse with the rest of the phylogenetic tree. It has a relatively small length, caused by an uncertainty about the tree topolgy. The density of the full posterior $\mu$ is of course difficult to visualize, but we can have a look at a small section. For this, consider the edges}
  e_2 &: \nsplit{Pika, Kangaroo rat}{Mouse, Squirrel, Guinea pig} \ , \quad \text{and}\\
  e_3 &: \nsplit{Pika, Squirrel}{Kangaroo rat, Mouse, Guinea pig} \ ,
\end{align*}
which are not compatible with $e_1$ and can replace it in the phylogenetic tree. Figure~\ref{fig:ex1-hist} shows histograms of lengths for the three edges, which can be interpreted as an estimate of a marginal posterior density. The histogram was generated by counting how often each of the edges appeared in the set of samples. We call the density marginal, because we did not consider a specific topology of the remaining tree. Hence, it does not reflect a single orthant of tree space. The estimate has positive support on all three edges. While Fig.~\ref{fig:ex1-hista} shows only a single mode, we clearly have a bimodality in Fig.~\ref{fig:ex1-histb}. Although the edge $e_1$ is present in the posterior expectation (Fig.~\ref{fig:ex1-mean}), its length is reduced due to the mass on $|e_2|$ and $|e_3|$. This correctly represents our uncertainty about the topology of the tree. For instance, an equal weight on all three edges would cause the posterior expectation to have a non-binary branching point.

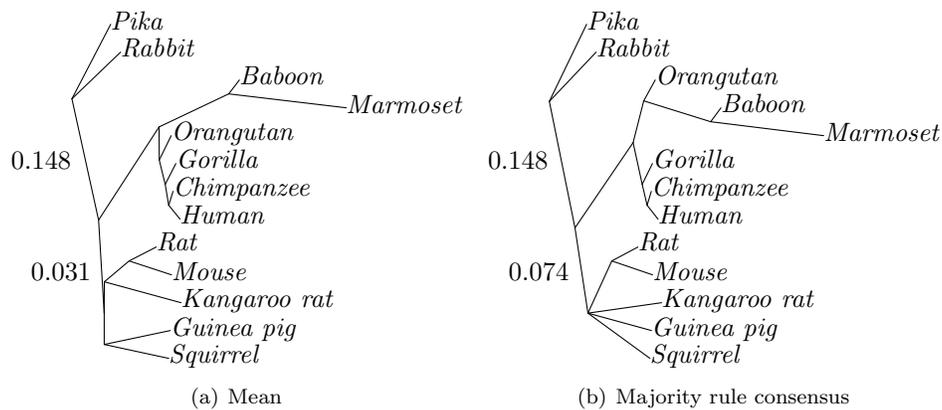
\begin{figure}[ht]
  \centering
  \subfigure[Mean]{
    \begin{tikzpicture}[>=latex,scale=0.63,x=1pt,y=1pt]
      \input{phylogenetic_example2.mean.tex}
    \end{tikzpicture}
  }
  \subfigure[Majority rule consensus]{
    \begin{tikzpicture}[>=latex,scale=0.63,x=1pt,y=1pt]
      \input{phylogenetic_example2.consensus.tex}
    \end{tikzpicture}
  }
  \caption{Example 2: Posterior \Frechet mean and majority rule consensus tree. The estimated posterior variance is $0.072$. Edge lengths are visualized as distances in the horizontal direction only.}
  \label{fig:ex2-mean}
\end{figure}

\begin{figure}[ht]
  \centering
  \begin{tikzpicture}[>=latex,scale=0.9,x=1pt,y=1pt]
    \input{phylogenetic_example2.histogram1.tex}
  \end{tikzpicture}
  \caption{Example 2: Marginal posterior density estimate of two edges. The posterior expectation is shown as a vertical line.}
  \label{fig:ex2-hist}
\end{figure}
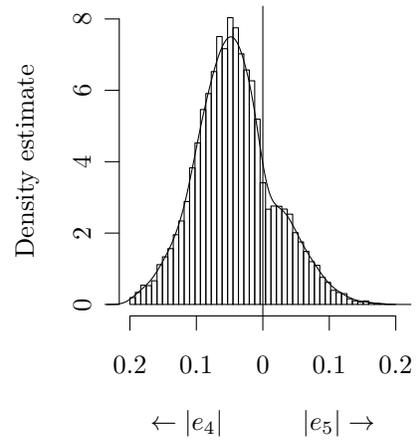

In a second example we increased the number of species to 13 and used 10 Markov chains in parallel. The approximated mean is shown in Fig.~\ref{fig:ex2-mean}. The edges that separate Kangaroo rat, Guinea pig, and Squirrel are of very short length, which shows that also in this example there is uncertainty about the exact topology of the tree. Figure~\ref{fig:ex2-hist} shows a marginal posterior estimate for the edges
\begin{align*}
  e_4 &: \nsplit{Guinea pig, Squirrel}{Pika, Rabbit, Kangaroo rat, Mouse, Rat, ... } \ , \quad \text{and}\\
  e_5 &: \nsplit{Kangaroo rat, Squirrel}{Pika, Rabbit, Guinea pig, Mouse, Rat, ... } \ ,
\end{align*}
which shows a strong bimodality. However, the interpretation of such marginal estimates is difficult because of the much richer structure of the full tree space. Figure~\ref{fig:ex2-mean} also shows the majority rule consensus tree. The topology of the tree is similar to the mean, but not identical. There is also no inner edge that separates Kangaroo rat, Guinea pig, and Squirrel, because no such edge appears in more than $50\%$ of the samples. Many software packages such as MrBayes also compute edge lengths for the majority rule consensus tree by considering only those edges that appear in the resulting tree. To compare edge lengths between methods, we consider the inner edge
\begin{equation*}
  \nsplit{Squirrel, Guinea pig, Kangaroo rat, Mouse, Rat}{Pika, Rabbit, ... }.
\end{equation*}
While this edge in the consensus tree has a length of $0.074$, the same edge in the \Frechet mean has only a length of $0.031$. The difference in length can be explained by the fact that the \Frechet mean considers all edges present in the posterior samples. On the other hand, the edge
\begin{equation*}
  \nsplit{Pika, Rabbit}{Baboon, Marmoset, Orangutan, Gorilla, ... }
\end{equation*}
has the same length in both trees because it appears in all posterior samples.



\section{Conclusion}
We have presented a statistical model for the inference of phylogenetic trees from multiple sequence alignments. The model is formulated on tree space by \citet{billera}, which is an Hadamard space and therefore allows to define the mean and median of a probability distribution. The approximation of posterior quantities is complicated and we have summarized some recent developments that contributed to this work. Despite the fact that the posterior distribution will in most cases be highly nontrivial, we demonstrated on a simple example that the mean or median as a point estimate can reflect the uncertainty about the topology of the tree. Current methods for phylogenetic tree inference that rely on MCMC sampling often compute a (majority rule) consensus tree. Such a tree can be justified from decision theoretic principles. However, we believe that we have proposed a more rigorous approach to solve this issue. Since our statistical model is defined on the BHV tree space with a given metric, its inherent properties become part of the model, which clearly has implications on posterior estimates. Certainly, a disadvantage of MCMC approximations in phylogenetic inference is that the number of different topologies grows super-exponentially with the number of leaves. The method might thus be inappropriate for the inference of large trees as the approximation of the posterior quantities might require too many samples.

We provide two freely available implementations at~\citeauthor{tfbayes} and~\citeauthor{trap}.



\FloatBarrier

\begin{figure}[ht]
  \centering
  \subfigure[Example 1]{
    \input{phylogenetic_example1.posterior.tex}
  }
  \subfigure[Example 2]{
    \input{phylogenetic_example2.posterior.tex}
  }
  \caption{Value of the log posterior density (not normalized) of MCMC samples.}
  \label{fig:ex-posterior}
\end{figure}
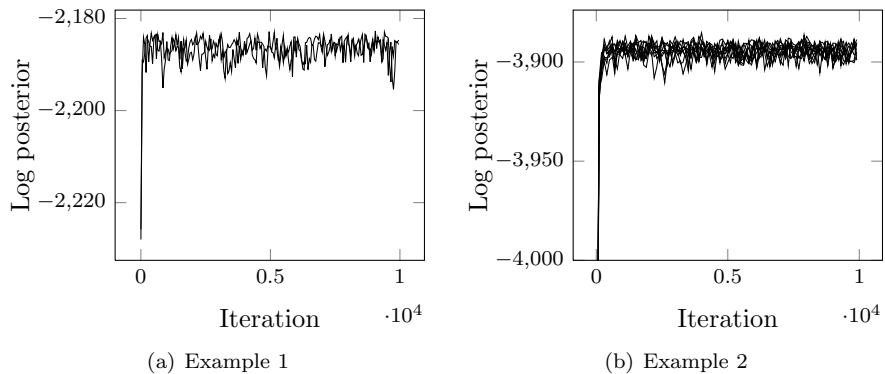

\FloatBarrier



\subsection*{Acknowledgements.}
We would like to thank Pierre-Yves Bourguignon, Stephan Poppe, and Johannes Schumacher for very valuable discussions. We are also extremely grateful to Ezra Miller and Megan Owen for their comments on the manuscript, which significantly improved the exposition.

\IfBiometrika{\mythanks}



\appendix
\input{phylogenetic_appendix1.tex}



\IfAmsart{%
  \bibliographystyle{plainnat}}
\IfBiometrika{%
  \bibliographystyle{biometrika}}
\bibliography{phylogenetic}



\end{document}

%% file: phylogenetic_triangle-b.pdf_tex
\begingroup%
  \makeatletter%
  \providecommand\color[2][]{%
    \errmessage{(Inkscape) Color is used for the text in Inkscape, but the package 'color.sty' is not loaded}%
    \renewcommand\color[2][]{}%
  }%
  \providecommand\transparent[1]{%
    \errmessage{(Inkscape) Transparency is used (non-zero) for the text in Inkscape, but the package 'transparent.sty' is not loaded}%
    \renewcommand\transparent[1]{}%
  }%
  \providecommand\rotatebox[2]{#2}%
  \ifx\svgwidth\undefined%
    \setlength{\unitlength}{186.68688965bp}%
    \ifx\svgscale\undefined%
      \relax%
    \else%
      \setlength{\unitlength}{\unitlength * \real{\svgscale}}%
    \fi%
  \else%
    \setlength{\unitlength}{\svgwidth}%
  \fi%
  \global\let\svgwidth\undefined%
  \global\let\svgscale\undefined%
  \makeatother%
  \begin{picture}(1,0.76075318)%
    \put(0,0){\includegraphics[width=\unitlength]{phylogenetic_triangle-b.pdf}}%
    \put(-0.00934051,0.02139275){\color[rgb]{0,0,0}\makebox(0,0)[lb]{\smash{$p$}}}%
    \put(0.57062581,0.68395346){\color[rgb]{0,0,0}\makebox(0,0)[lb]{\smash{$r$}}}%
    \put(0.92540987,0.25990939){\color[rgb]{0,0,0}\makebox(0,0)[lb]{\smash{$q$}}}%
  \end{picture}%
\endgroup%

%% file: phylogenetic_triangle-a.pdf_tex
\begingroup%
  \makeatletter%
  \providecommand\color[2][]{%
    \errmessage{(Inkscape) Color is used for the text in Inkscape, but the package 'color.sty' is not loaded}%
    \renewcommand\color[2][]{}%
  }%
  \providecommand\transparent[1]{%
    \errmessage{(Inkscape) Transparency is used (non-zero) for the text in Inkscape, but the package 'transparent.sty' is not loaded}%
    \renewcommand\transparent[1]{}%
  }%
  \providecommand\rotatebox[2]{#2}%
  \ifx\svgwidth\undefined%
    \setlength{\unitlength}{186.68688965bp}%
    \ifx\svgscale\undefined%
      \relax%
    \else%
      \setlength{\unitlength}{\unitlength * \real{\svgscale}}%
    \fi%
  \else%
    \setlength{\unitlength}{\svgwidth}%
  \fi%
  \global\let\svgwidth\undefined%
  \global\let\svgscale\undefined%
  \makeatother%
  \begin{picture}(1,0.76075318)%
    \put(0,0){\includegraphics[width=\unitlength]{phylogenetic_triangle-a.pdf}}%
    \put(-0.00934051,0.02139275){\color[rgb]{0,0,0}\makebox(0,0)[lb]{\smash{$\bar{p}$}}}%
    \put(0.57062581,0.68395346){\color[rgb]{0,0,0}\makebox(0,0)[lb]{\smash{$\bar{r}$}}}%
    \put(0.92540987,0.25990939){\color[rgb]{0,0,0}\makebox(0,0)[lb]{\smash{$\bar{q}$}}}%
  \end{picture}%
\endgroup%

%% file: phylogenetic_mcmc-moves-a.tex
\pgfsetcolor{black}
  \draw [very thick] (95bp,203.96bp) .. controls (95bp,187.27bp) and (95bp,165bp)  .. (95bp,165bp);
  \draw [very thick] (95bp,165bp) .. controls (95bp,165bp) and (63bp,92bp)  .. (63bp,92bp);
  \definecolor{strokecol}{rgb}{0.0,0.0,0.0};
  \pgfsetstrokecolor{strokecol}
  \draw (89.5bp,137bp) node[left=0.4cm] {$e$};
  \draw [very thick] (63bp,92bp) .. controls (63bp,92bp) and (46.56bp,58.206bp)  .. (35.899bp,36.292bp);
  \draw [very thick] (95bp,165bp) .. controls (95bp,165bp) and (100.94bp,131.66bp)  .. (104.79bp,110.04bp);
  \draw [very thick] (63bp,92bp) .. controls (63bp,92bp) and (79.44bp,58.206bp)  .. (90.101bp,36.292bp);
\begin{scope}
  \definecolor{strokecol}{rgb}{0.0,0.0,0.0};
  \pgfsetstrokecolor{strokecol}
  \draw (108bp,92bp) node {3};
\end{scope}
\begin{scope}
  \definecolor{strokecol}{rgb}{0.0,0.0,0.0};
  \pgfsetstrokecolor{strokecol}
  \draw (99bp,18bp) node {2};
\end{scope}
\begin{scope}
  \definecolor{strokecol}{rgb}{0.0,0.0,0.0};
  \pgfsetstrokecolor{strokecol}
  \draw (27bp,18bp) node {1};
\end{scope}
\begin{scope}
  \definecolor{strokecol}{rgb}{0.0,0.0,0.0};
  \pgfsetstrokecolor{strokecol}
  \draw (95bp,222bp) node {0};
\end{scope}

%% file: phylogenetic_mcmc-moves-b.tex
\pgfsetcolor{black}
  \draw [very thick] (94bp,203.96bp) .. controls (94bp,187.27bp) and (94bp,165bp)  .. (94bp,165bp);
  \draw [very thick] (94bp,165bp) .. controls (94bp,165bp) and (63bp,92bp)  .. (63bp,92bp);
  \definecolor{strokecol}{rgb}{0.0,0.0,0.0};
  \pgfsetstrokecolor{strokecol}
  \draw (87bp,137bp) node[left=0.4cm] {$e'$};
  \draw [very thick] (63bp,92bp) .. controls (63bp,92bp) and (46.56bp,58.206bp)  .. (35.899bp,36.292bp);
  \draw [very thick] (94bp,165bp) .. controls (94bp,165bp) and (100.39bp,131.66bp)  .. (104.54bp,110.04bp);
  \draw [very thick] (63bp,92bp) .. controls (63bp,92bp) and (79.44bp,58.206bp)  .. (90.101bp,36.292bp);
\begin{scope}
  \definecolor{strokecol}{rgb}{0.0,0.0,0.0};
  \pgfsetstrokecolor{strokecol}
  \draw (108bp,92bp) node {1};
\end{scope}
\begin{scope}
  \definecolor{strokecol}{rgb}{0.0,0.0,0.0};
  \pgfsetstrokecolor{strokecol}
  \draw (99bp,18bp) node {2};
\end{scope}
\begin{scope}
  \definecolor{strokecol}{rgb}{0.0,0.0,0.0};
  \pgfsetstrokecolor{strokecol}
  \draw (27bp,18bp) node {3};
\end{scope}
\begin{scope}
  \definecolor{strokecol}{rgb}{0.0,0.0,0.0};
  \pgfsetstrokecolor{strokecol}
  \draw (94bp,222bp) node {0};
\end{scope}

%% file: phylogenetic_mcmc-moves-c.tex
\pgfsetcolor{black}
  \draw [very thick] (94bp,203.96bp) .. controls (94bp,187.27bp) and (94bp,165bp)  .. (94bp,165bp);
  \draw [very thick] (94bp,165bp) .. controls (94bp,165bp) and (63bp,92bp)  .. (63bp,92bp);
  \definecolor{strokecol}{rgb}{0.0,0.0,0.0};
  \pgfsetstrokecolor{strokecol}
  \draw (87bp,137bp) node[left=0.4cm] {$e'$};
  \draw [very thick] (63bp,92bp) .. controls (63bp,92bp) and (46.56bp,58.206bp)  .. (35.899bp,36.292bp);
  \draw [very thick] (94bp,165bp) .. controls (94bp,165bp) and (100.39bp,131.66bp)  .. (104.54bp,110.04bp);
  \draw [very thick] (63bp,92bp) .. controls (63bp,92bp) and (79.44bp,58.206bp)  .. (90.101bp,36.292bp);
\begin{scope}
  \definecolor{strokecol}{rgb}{0.0,0.0,0.0};
  \pgfsetstrokecolor{strokecol}
  \draw (108bp,92bp) node {2};
\end{scope}
\begin{scope}
  \definecolor{strokecol}{rgb}{0.0,0.0,0.0};
  \pgfsetstrokecolor{strokecol}
  \draw (99bp,18bp) node {3};
\end{scope}
\begin{scope}
  \definecolor{strokecol}{rgb}{0.0,0.0,0.0};
  \pgfsetstrokecolor{strokecol}
  \draw (27bp,18bp) node {1};
\end{scope}
\begin{scope}
  \definecolor{strokecol}{rgb}{0.0,0.0,0.0};
  \pgfsetstrokecolor{strokecol}
  \draw (94bp,222bp) node {0};
\end{scope}

%% file: phylogenetic_example1.posterior.tex
\begin{tikzpicture}
  \pgfplotsset{
    tick label style={font=\footnotesize}
  }
  \begin{axis}[
      width=.45\textwidth,
      xlabel={Iteration},
      ylabel={Log posterior},
    ]

    \addplot[color=black] table[x=x,y=y1] {phylogenetic_example1.posterior.dat};
    \addplot[color=black] table[x=x,y=y2] {phylogenetic_example1.posterior.dat};

  \end{axis}
\end{tikzpicture}

%% file: phylogenetic_example2.posterior.tex
\begin{tikzpicture}
  \pgfplotsset{
    tick label style={font=\footnotesize}
  }
  \begin{axis}[
      width=.45\textwidth,
      xlabel={Iteration},
      ylabel={Log posterior},
      ymin=-4000
    ]

    \addplot[color=black] table[x=x,y=y1] {phylogenetic_example2.posterior.dat};
    \addplot[color=black] table[x=x,y=y2] {phylogenetic_example2.posterior.dat};
    \addplot[color=black] table[x=x,y=y3] {phylogenetic_example2.posterior.dat};
    \addplot[color=black] table[x=x,y=y4] {phylogenetic_example2.posterior.dat};
    \addplot[color=black] table[x=x,y=y5] {phylogenetic_example2.posterior.dat};
    \addplot[color=black] table[x=x,y=y6] {phylogenetic_example2.posterior.dat};
    \addplot[color=black] table[x=x,y=y7] {phylogenetic_example2.posterior.dat};
    \addplot[color=black] table[x=x,y=y8] {phylogenetic_example2.posterior.dat};
    \addplot[color=black] table[x=x,y=y9] {phylogenetic_example2.posterior.dat};
    \addplot[color=black] table[x=x,y=y10] {phylogenetic_example2.posterior.dat};

  \end{axis}
\end{tikzpicture}

%% file: phylogenetic_example1.mean.tex
\definecolor[named]{drawColor}{rgb}{0.00,0.00,0.00}
\definecolor[named]{fillColor}{rgb}{1.00,1.00,1.00}
\fill[color=fillColor,] (0,0) rectangle (144.54,144.54);
\begin{scope}

\definecolor[named]{drawColor}{rgb}{0.00,0.00,0.00}

\draw[color=drawColor,line cap=round,line join=round,] (  5.35, 97.36) -- (  9.25, 47.18);

\draw[color=drawColor,line cap=round,line join=round,] (  9.25, 47.18) -- ( 26.61, 22.08);

\draw[color=drawColor,line cap=round,line join=round,] ( 26.61, 22.08) -- (104.16,  5.35);

\draw[color=drawColor,line cap=round,line join=round,] ( 26.61, 22.08) -- (102.98, 38.81);

\draw[color=drawColor,line cap=round,line join=round,] (  9.25, 47.18) -- ( 91.60, 72.27);

\draw[color=drawColor,line cap=round,line join=round,] (  5.35, 97.36) -- ( 73.45,105.73);

\draw[color=drawColor,line cap=round,line join=round,] (  5.35, 97.36) -- ( 78.39,139.19);

\node[color=drawColor,anchor=base west,inner sep=0pt, outer sep=0pt, scale=  1.00] at (104.16,  2.88) {\itshape Squirrel};

\node[color=drawColor,anchor=base west,inner sep=0pt, outer sep=0pt, scale=  1.00] at (102.98, 35.37) {\itshape Guinea pig};

\node[color=drawColor,anchor=base west,inner sep=0pt, outer sep=0pt, scale=  1.00] at ( 91.60, 69.80) {\itshape Kangaroo rat};

\node[color=drawColor,anchor=base west,inner sep=0pt, outer sep=0pt, scale=  1.00] at ( 73.45,102.52) {\itshape Mouse};

\node[color=drawColor,anchor=base west,inner sep=0pt, outer sep=0pt, scale=  1.00] at ( 78.39,135.74) {\itshape Pika};
\end{scope}

%% file: phylogenetic_example1.histogram1.tex
\definecolor[named]{drawColor}{rgb}{0.00,0.00,0.00}
\definecolor[named]{fillColor}{rgb}{1.00,1.00,1.00}
\fill[color=fillColor,] (0,0) rectangle (180.67,180.67);
\begin{scope}

\definecolor[named]{drawColor}{rgb}{0.00,0.00,0.00}

\node[color=drawColor,anchor=base,inner sep=0pt, outer sep=0pt, scale=  1.00] at (108.34,  2.40) {$\leftarrow|e_1|$\quad\quad\quad$|e_2|\rightarrow$};

\node[rotate= 90.00,color=drawColor,anchor=base,inner sep=0pt, outer sep=0pt, scale=  1.00] at ( 12.00,115.54) {Density estimate};
\end{scope}
\begin{scope}

\definecolor[named]{drawColor}{rgb}{0.00,0.00,0.00}

\draw[color=drawColor,line cap=round,line join=round,] ( 52.47, 55.23) rectangle ( 54.75, 55.74);

\draw[color=drawColor,line cap=round,line join=round,] ( 54.75, 55.23) rectangle ( 57.03, 57.39);

\draw[color=drawColor,line cap=round,line join=round,] ( 57.03, 55.23) rectangle ( 59.31, 57.78);

\draw[color=drawColor,line cap=round,line join=round,] ( 59.31, 55.23) rectangle ( 61.59, 64.80);

\draw[color=drawColor,line cap=round,line join=round,] ( 61.59, 55.23) rectangle ( 63.87, 66.46);

\draw[color=drawColor,line cap=round,line join=round,] ( 63.87, 55.23) rectangle ( 66.15, 70.80);

\draw[color=drawColor,line cap=round,line join=round,] ( 66.15, 55.23) rectangle ( 68.43, 75.14);

\draw[color=drawColor,line cap=round,line join=round,] ( 68.43, 55.23) rectangle ( 70.71, 76.92);

\draw[color=drawColor,line cap=round,line join=round,] ( 70.71, 55.23) rectangle ( 72.99, 79.86);

\draw[color=drawColor,line cap=round,line join=round,] ( 72.99, 55.23) rectangle ( 75.27, 93.39);

\draw[color=drawColor,line cap=round,line join=round,] ( 75.27, 55.23) rectangle ( 77.55, 96.97);

\draw[color=drawColor,line cap=round,line join=round,] ( 77.55, 55.23) rectangle ( 79.83,111.01);

\draw[color=drawColor,line cap=round,line join=round,] ( 79.83, 55.23) rectangle ( 82.11,130.15);

\draw[color=drawColor,line cap=round,line join=round,] ( 82.11, 55.23) rectangle ( 84.39,139.09);

\draw[color=drawColor,line cap=round,line join=round,] ( 84.39, 55.23) rectangle ( 86.67,150.45);

\draw[color=drawColor,line cap=round,line join=round,] ( 86.67, 55.23) rectangle ( 88.95,163.47);

\draw[color=drawColor,line cap=round,line join=round,] ( 88.95, 55.23) rectangle ( 91.24,170.87);

\draw[color=drawColor,line cap=round,line join=round,] ( 91.24, 55.23) rectangle ( 93.52,169.98);

\draw[color=drawColor,line cap=round,line join=round,] ( 93.52, 55.23) rectangle ( 95.80,175.85);

\draw[color=drawColor,line cap=round,line join=round,] ( 95.80, 55.23) rectangle ( 98.08,163.47);

\draw[color=drawColor,line cap=round,line join=round,] ( 98.08, 55.23) rectangle (100.36,143.05);

\draw[color=drawColor,line cap=round,line join=round,] (100.36, 55.23) rectangle (102.64,129.77);

\draw[color=drawColor,line cap=round,line join=round,] (102.64, 55.23) rectangle (104.92,105.77);

\draw[color=drawColor,line cap=round,line join=round,] (104.92, 55.23) rectangle (107.20, 90.07);

\draw[color=drawColor,line cap=round,line join=round,] (107.20, 55.23) rectangle (109.48, 82.29);

\draw[color=drawColor,line cap=round,line join=round,] (109.48, 55.23) rectangle (111.76, 78.07);

\draw[color=drawColor,line cap=round,line join=round,] (111.76, 55.23) rectangle (114.04, 70.16);

\draw[color=drawColor,line cap=round,line join=round,] (114.04, 55.23) rectangle (116.32, 65.95);

\draw[color=drawColor,line cap=round,line join=round,] (116.32, 55.23) rectangle (118.60, 66.59);

\draw[color=drawColor,line cap=round,line join=round,] (118.60, 55.23) rectangle (120.88, 61.10);

\draw[color=drawColor,line cap=round,line join=round,] (120.88, 55.23) rectangle (123.16, 61.61);

\draw[color=drawColor,line cap=round,line join=round,] (123.16, 55.23) rectangle (125.44, 59.69);

\draw[color=drawColor,line cap=round,line join=round,] (125.44, 55.23) rectangle (127.72, 56.76);

\draw[color=drawColor,line cap=round,line join=round,] (127.72, 55.23) rectangle (130.00, 57.39);

\draw[color=drawColor,line cap=round,line join=round,] (130.00, 55.23) rectangle (132.28, 56.76);

\draw[color=drawColor,line cap=round,line join=round,] (132.28, 55.23) rectangle (134.56, 55.48);

\draw[color=drawColor,line cap=round,line join=round,] (134.56, 55.23) rectangle (136.84, 55.61);

\draw[color=drawColor,line cap=round,line join=round,] (136.84, 55.23) rectangle (139.12, 55.61);

\draw[color=drawColor,line cap=round,line join=round,] (139.12, 55.23) rectangle (141.40, 55.23);

\draw[color=drawColor,line cap=round,line join=round,] (141.40, 55.23) rectangle (143.68, 55.35);

\draw[color=drawColor,line cap=round,line join=round,] (143.68, 55.23) rectangle (145.96, 55.35);

\draw[color=drawColor,line cap=round,line join=round,] (145.96, 55.23) rectangle (148.24, 55.23);

\draw[color=drawColor,line cap=round,line join=round,] (148.24, 55.23) rectangle (150.52, 55.23);

\draw[color=drawColor,line cap=round,line join=round,] (150.52, 55.23) rectangle (152.80, 55.23);

\draw[color=drawColor,line cap=round,line join=round,] (152.80, 55.23) rectangle (155.08, 55.23);

\draw[color=drawColor,line cap=round,line join=round,] (155.08, 55.23) rectangle (157.36, 55.23);

\draw[color=drawColor,line cap=round,line join=round,] (157.36, 55.23) rectangle (159.64, 55.23);

\draw[color=drawColor,line cap=round,line join=round,] (159.64, 55.23) rectangle (161.93, 55.23);

\draw[color=drawColor,line cap=round,line join=round,] (161.93, 55.23) rectangle (164.21, 55.23);

\draw[color=drawColor,line cap=round,line join=round,] ( 43.88, 55.23) --
	( 44.09, 55.23) --
	( 44.31, 55.23) --
	( 44.52, 55.23) --
	( 44.74, 55.23) --
	( 44.95, 55.23) --
	( 45.16, 55.23) --
	( 45.38, 55.23) --
	( 45.59, 55.24) --
	( 45.81, 55.24) --
	( 46.02, 55.24) --
	( 46.23, 55.24) --
	( 46.45, 55.25) --
	( 46.66, 55.25) --
	( 46.87, 55.26) --
	( 47.09, 55.26) --
	( 47.30, 55.27) --
	( 47.52, 55.27) --
	( 47.73, 55.28) --
	( 47.94, 55.29) --
	( 48.16, 55.30) --
	( 48.37, 55.31) --
	( 48.59, 55.33) --
	( 48.80, 55.34) --
	( 49.01, 55.36) --
	( 49.23, 55.37) --
	( 49.44, 55.39) --
	( 49.66, 55.42) --
	( 49.87, 55.44) --
	( 50.08, 55.47) --
	( 50.30, 55.50) --
	( 50.51, 55.53) --
	( 50.72, 55.56) --
	( 50.94, 55.60) --
	( 51.15, 55.64) --
	( 51.37, 55.69) --
	( 51.58, 55.73) --
	( 51.79, 55.79) --
	( 52.01, 55.84) --
	( 52.22, 55.90) --
	( 52.44, 55.97) --
	( 52.65, 56.04) --
	( 52.86, 56.11) --
	( 53.08, 56.19) --
	( 53.29, 56.28) --
	( 53.50, 56.37) --
	( 53.72, 56.46) --
	( 53.93, 56.57) --
	( 54.15, 56.68) --
	( 54.36, 56.79) --
	( 54.57, 56.92) --
	( 54.79, 57.05) --
	( 55.00, 57.19) --
	( 55.22, 57.34) --
	( 55.43, 57.49) --
	( 55.64, 57.66) --
	( 55.86, 57.83) --
	( 56.07, 58.01) --
	( 56.28, 58.21) --
	( 56.50, 58.41) --
	( 56.71, 58.62) --
	( 56.93, 58.84) --
	( 57.14, 59.07) --
	( 57.35, 59.30) --
	( 57.57, 59.55) --
	( 57.78, 59.81) --
	( 58.00, 60.07) --
	( 58.21, 60.35) --
	( 58.42, 60.63) --
	( 58.64, 60.92) --
	( 58.85, 61.21) --
	( 59.06, 61.52) --
	( 59.28, 61.82) --
	( 59.49, 62.14) --
	( 59.71, 62.46) --
	( 59.92, 62.79) --
	( 60.13, 63.12) --
	( 60.35, 63.45) --
	( 60.56, 63.78) --
	( 60.78, 64.12) --
	( 60.99, 64.46) --
	( 61.20, 64.81) --
	( 61.42, 65.15) --
	( 61.63, 65.49) --
	( 61.85, 65.84) --
	( 62.06, 66.18) --
	( 62.27, 66.52) --
	( 62.49, 66.86) --
	( 62.70, 67.20) --
	( 62.91, 67.54) --
	( 63.13, 67.88) --
	( 63.34, 68.21) --
	( 63.56, 68.54) --
	( 63.77, 68.87) --
	( 63.98, 69.20) --
	( 64.20, 69.53) --
	( 64.41, 69.86) --
	( 64.63, 70.18) --
	( 64.84, 70.51) --
	( 65.05, 70.83) --
	( 65.27, 71.16) --
	( 65.48, 71.48) --
	( 65.69, 71.81) --
	( 65.91, 72.13) --
	( 66.12, 72.46) --
	( 66.34, 72.79) --
	( 66.55, 73.13) --
	( 66.76, 73.47) --
	( 66.98, 73.81) --
	( 67.19, 74.16) --
	( 67.41, 74.51) --
	( 67.62, 74.87) --
	( 67.83, 75.24) --
	( 68.05, 75.62) --
	( 68.26, 76.01) --
	( 68.47, 76.41) --
	( 68.69, 76.81) --
	( 68.90, 77.23) --
	( 69.12, 77.67) --
	( 69.33, 78.11) --
	( 69.54, 78.57) --
	( 69.76, 79.04) --
	( 69.97, 79.53) --
	( 70.19, 80.03) --
	( 70.40, 80.54) --
	( 70.61, 81.07) --
	( 70.83, 81.61) --
	( 71.04, 82.17) --
	( 71.26, 82.75) --
	( 71.47, 83.34) --
	( 71.68, 83.95) --
	( 71.90, 84.57) --
	( 72.11, 85.20) --
	( 72.32, 85.85) --
	( 72.54, 86.52) --
	( 72.75, 87.19) --
	( 72.97, 87.89) --
	( 73.18, 88.59) --
	( 73.39, 89.32) --
	( 73.61, 90.05) --
	( 73.82, 90.81) --
	( 74.04, 91.57) --
	( 74.25, 92.36) --
	( 74.46, 93.15) --
	( 74.68, 93.97) --
	( 74.89, 94.81) --
	( 75.10, 95.66) --
	( 75.32, 96.53) --
	( 75.53, 97.42) --
	( 75.75, 98.33) --
	( 75.96, 99.27) --
	( 76.17,100.22) --
	( 76.39,101.20) --
	( 76.60,102.19) --
	( 76.82,103.21) --
	( 77.03,104.26) --
	( 77.24,105.32) --
	( 77.46,106.41) --
	( 77.67,107.52) --
	( 77.88,108.64) --
	( 78.10,109.79) --
	( 78.31,110.96) --
	( 78.53,112.14) --
	( 78.74,113.34) --
	( 78.95,114.55) --
	( 79.17,115.77) --
	( 79.38,117.01) --
	( 79.60,118.25) --
	( 79.81,119.49) --
	( 80.02,120.74) --
	( 80.24,121.99) --
	( 80.45,123.24) --
	( 80.67,124.49) --
	( 80.88,125.73) --
	( 81.09,126.97) --
	( 81.31,128.20) --
	( 81.52,129.42) --
	( 81.73,130.64) --
	( 81.95,131.84) --
	( 82.16,133.03) --
	( 82.38,134.22) --
	( 82.59,135.38) --
	( 82.80,136.54) --
	( 83.02,137.68) --
	( 83.23,138.81) --
	( 83.45,139.93) --
	( 83.66,141.04) --
	( 83.87,142.13) --
	( 84.09,143.20) --
	( 84.30,144.27) --
	( 84.51,145.33) --
	( 84.73,146.37) --
	( 84.94,147.39) --
	( 85.16,148.41) --
	( 85.37,149.41) --
	( 85.58,150.39) --
	( 85.80,151.37) --
	( 86.01,152.32) --
	( 86.23,153.26) --
	( 86.44,154.18) --
	( 86.65,155.09) --
	( 86.87,155.97) --
	( 87.08,156.84) --
	( 87.29,157.68) --
	( 87.51,158.50) --
	( 87.72,159.30) --
	( 87.94,160.08) --
	( 88.15,160.83) --
	( 88.36,161.54) --
	( 88.58,162.24) --
	( 88.79,162.90) --
	( 89.01,163.53) --
	( 89.22,164.14) --
	( 89.43,164.71) --
	( 89.65,165.25) --
	( 89.86,165.76) --
	( 90.07,166.23) --
	( 90.29,166.68) --
	( 90.50,167.09) --
	( 90.72,167.45) --
	( 90.93,167.79) --
	( 91.14,168.09) --
	( 91.36,168.35) --
	( 91.57,168.58) --
	( 91.79,168.76) --
	( 92.00,168.90) --
	( 92.21,169.00) --
	( 92.43,169.05) --
	( 92.64,169.07) --
	( 92.86,169.04) --
	( 93.07,168.96) --
	( 93.28,168.82) --
	( 93.50,168.64) --
	( 93.71,168.42) --
	( 93.92,168.14) --
	( 94.14,167.79) --
	( 94.35,167.39) --
	( 94.57,166.94) --
	( 94.78,166.43) --
	( 94.99,165.86) --
	( 95.21,165.23) --
	( 95.42,164.54) --
	( 95.64,163.79) --
	( 95.85,162.98) --
	( 96.06,162.12) --
	( 96.28,161.20) --
	( 96.49,160.21) --
	( 96.70,159.17) --
	( 96.92,158.08) --
	( 97.13,156.94) --
	( 97.35,155.76) --
	( 97.56,154.52) --
	( 97.77,153.23) --
	( 97.99,151.90) --
	( 98.20,150.54) --
	( 98.42,149.14) --
	( 98.63,147.70) --
	( 98.84,146.23) --
	( 99.06,144.74) --
	( 99.27,143.21) --
	( 99.48,141.67) --
	( 99.70,140.11) --
	( 99.91,138.52) --
	(100.13,136.92) --
	(100.34,135.31) --
	(100.55,133.69) --
	(100.77,132.07) --
	(100.98,130.43) --
	(101.20,128.80) --
	(101.41,127.16) --
	(101.62,125.53) --
	(101.84,123.90) --
	(102.05,122.27) --
	(102.27,120.66) --
	(102.48,119.05) --
	(102.69,117.46) --
	(102.91,115.88) --
	(103.12,114.32) --
	(103.33,112.78) --
	(103.55,111.26) --
	(103.76,109.76) --
	(103.98,108.29) --
	(104.19,106.84) --
	(104.40,105.42) --
	(104.62,104.03) --
	(104.83,102.67) --
	(105.05,101.35) --
	(105.26,100.05) --
	(105.47, 98.79) --
	(105.69, 97.57) --
	(105.90, 96.38) --
	(106.11, 95.22) --
	(106.33, 94.10) --
	(106.54, 93.01) --
	(106.76, 91.97) --
	(106.97, 90.95) --
	(107.18, 89.97) --
	(107.40, 89.02) --
	(107.61, 88.10) --
	(107.83, 87.22) --
	(108.04, 86.37) --
	(108.25, 85.54) --
	(108.47, 84.74) --
	(108.68, 83.97) --
	(108.89, 83.22) --
	(109.11, 82.50) --
	(109.32, 81.80) --
	(109.54, 81.11) --
	(109.75, 80.45) --
	(109.96, 79.80) --
	(110.18, 79.18) --
	(110.39, 78.57) --
	(110.61, 77.97) --
	(110.82, 77.39) --
	(111.03, 76.82) --
	(111.25, 76.27) --
	(111.46, 75.73) --
	(111.67, 75.20) --
	(111.89, 74.68) --
	(112.10, 74.17) --
	(112.32, 73.68) --
	(112.53, 73.19) --
	(112.74, 72.72) --
	(112.96, 72.26) --
	(113.17, 71.81) --
	(113.39, 71.37) --
	(113.60, 70.95) --
	(113.81, 70.53) --
	(114.03, 70.13) --
	(114.24, 69.74) --
	(114.46, 69.36) --
	(114.67, 68.99) --
	(114.88, 68.64) --
	(115.10, 68.29) --
	(115.31, 67.96) --
	(115.52, 67.64) --
	(115.74, 67.32) --
	(115.95, 67.02) --
	(116.17, 66.73) --
	(116.38, 66.45) --
	(116.59, 66.18) --
	(116.81, 65.92) --
	(117.02, 65.66) --
	(117.24, 65.42) --
	(117.45, 65.18) --
	(117.66, 64.94) --
	(117.88, 64.72) --
	(118.09, 64.50) --
	(118.30, 64.28) --
	(118.52, 64.07) --
	(118.73, 63.87) --
	(118.95, 63.67) --
	(119.16, 63.47) --
	(119.37, 63.28) --
	(119.59, 63.09) --
	(119.80, 62.91) --
	(120.02, 62.72) --
	(120.23, 62.55) --
	(120.44, 62.37) --
	(120.66, 62.20) --
	(120.87, 62.03) --
	(121.08, 61.86) --
	(121.30, 61.69) --
	(121.51, 61.53) --
	(121.73, 61.37) --
	(121.94, 61.21) --
	(122.15, 61.05) --
	(122.37, 60.89) --
	(122.58, 60.73) --
	(122.80, 60.58) --
	(123.01, 60.43) --
	(123.22, 60.28) --
	(123.44, 60.13) --
	(123.65, 59.99) --
	(123.87, 59.84) --
	(124.08, 59.70) --
	(124.29, 59.56) --
	(124.51, 59.42) --
	(124.72, 59.28) --
	(124.93, 59.15) --
	(125.15, 59.02) --
	(125.36, 58.89) --
	(125.58, 58.76) --
	(125.79, 58.64) --
	(126.00, 58.52) --
	(126.22, 58.41) --
	(126.43, 58.29) --
	(126.65, 58.18) --
	(126.86, 58.07) --
	(127.07, 57.97) --
	(127.29, 57.87) --
	(127.50, 57.77) --
	(127.71, 57.68) --
	(127.93, 57.59) --
	(128.14, 57.50) --
	(128.36, 57.41) --
	(128.57, 57.33) --
	(128.78, 57.25) --
	(129.00, 57.17) --
	(129.21, 57.10) --
	(129.43, 57.03) --
	(129.64, 56.95) --
	(129.85, 56.88) --
	(130.07, 56.82) --
	(130.28, 56.75) --
	(130.49, 56.69) --
	(130.71, 56.63) --
	(130.92, 56.57) --
	(131.14, 56.51) --
	(131.35, 56.45) --
	(131.56, 56.40) --
	(131.78, 56.34) --
	(131.99, 56.29) --
	(132.21, 56.24) --
	(132.42, 56.19) --
	(132.63, 56.14) --
	(132.85, 56.10) --
	(133.06, 56.06) --
	(133.27, 56.02) --
	(133.49, 55.98) --
	(133.70, 55.94) --
	(133.92, 55.90) --
	(134.13, 55.87) --
	(134.34, 55.84) --
	(134.56, 55.81) --
	(134.77, 55.78) --
	(134.99, 55.75) --
	(135.20, 55.73) --
	(135.41, 55.70) --
	(135.63, 55.68) --
	(135.84, 55.66) --
	(136.06, 55.64) --
	(136.27, 55.62) --
	(136.48, 55.60) --
	(136.70, 55.58) --
	(136.91, 55.57) --
	(137.12, 55.55) --
	(137.34, 55.53) --
	(137.55, 55.52) --
	(137.77, 55.51) --
	(137.98, 55.49) --
	(138.19, 55.48) --
	(138.41, 55.47) --
	(138.62, 55.46) --
	(138.84, 55.45) --
	(139.05, 55.44) --
	(139.26, 55.43) --
	(139.48, 55.42) --
	(139.69, 55.41) --
	(139.90, 55.41) --
	(140.12, 55.40) --
	(140.33, 55.39) --
	(140.55, 55.39) --
	(140.76, 55.38) --
	(140.97, 55.37) --
	(141.19, 55.37) --
	(141.40, 55.36) --
	(141.62, 55.36) --
	(141.83, 55.36) --
	(142.04, 55.35) --
	(142.26, 55.35) --
	(142.47, 55.34) --
	(142.68, 55.34) --
	(142.90, 55.34) --
	(143.11, 55.33) --
	(143.33, 55.33) --
	(143.54, 55.33) --
	(143.75, 55.32) --
	(143.97, 55.32) --
	(144.18, 55.32) --
	(144.40, 55.31) --
	(144.61, 55.31) --
	(144.82, 55.31) --
	(145.04, 55.30) --
	(145.25, 55.30) --
	(145.47, 55.29) --
	(145.68, 55.29) --
	(145.89, 55.29) --
	(146.11, 55.28) --
	(146.32, 55.28) --
	(146.53, 55.28) --
	(146.75, 55.27) --
	(146.96, 55.27) --
	(147.18, 55.27) --
	(147.39, 55.26) --
	(147.60, 55.26) --
	(147.82, 55.26) --
	(148.03, 55.25) --
	(148.25, 55.25) --
	(148.46, 55.25) --
	(148.67, 55.25) --
	(148.89, 55.24) --
	(149.10, 55.24) --
	(149.31, 55.24) --
	(149.53, 55.24) --
	(149.74, 55.24) --
	(149.96, 55.23) --
	(150.17, 55.23) --
	(150.38, 55.23) --
	(150.60, 55.23) --
	(150.81, 55.23) --
	(151.03, 55.23) --
	(151.24, 55.23) --
	(151.45, 55.23) --
	(151.67, 55.23) --
	(151.88, 55.23) --
	(152.09, 55.23) --
	(152.31, 55.23) --
	(152.52, 55.23) --
	(152.74, 55.23) --
	(152.95, 55.23) --
	(153.16, 55.23);
\end{scope}
\begin{scope}

\definecolor[named]{drawColor}{rgb}{0.00,0.00,0.00}

\draw[color=drawColor,line cap=round,line join=round,] ( 52.47, 50.40) -- (164.21, 50.40);

\draw[color=drawColor,line cap=round,line join=round,] ( 52.47, 50.40) -- ( 52.47, 44.40);

\draw[color=drawColor,line cap=round,line join=round,] ( 80.40, 50.40) -- ( 80.40, 44.40);

\draw[color=drawColor,line cap=round,line join=round,] (108.34, 50.40) -- (108.34, 44.40);

\draw[color=drawColor,line cap=round,line join=round,] (136.27, 50.40) -- (136.27, 44.40);

\draw[color=drawColor,line cap=round,line join=round,] (164.21, 50.40) -- (164.21, 44.40);

\node[color=drawColor,anchor=base,inner sep=0pt, outer sep=0pt, scale=  1.00] at ( 52.47, 26.40) {0.2};

\node[color=drawColor,anchor=base,inner sep=0pt, outer sep=0pt, scale=  1.00] at ( 80.40, 26.40) {0.1};

\node[color=drawColor,anchor=base,inner sep=0pt, outer sep=0pt, scale=  1.00] at (108.34, 26.40) {0};

\node[color=drawColor,anchor=base,inner sep=0pt, outer sep=0pt, scale=  1.00] at (136.27, 26.40) {0.1};

\node[color=drawColor,anchor=base,inner sep=0pt, outer sep=0pt, scale=  1.00] at (164.21, 26.40) {0.2};

\draw[color=drawColor,line cap=round,line join=round,] ( 48.00, 55.23) -- ( 48.00,171.39);

\draw[color=drawColor,line cap=round,line join=round,] ( 48.00, 55.23) -- ( 42.00, 55.23);

\draw[color=drawColor,line cap=round,line join=round,] ( 48.00, 78.46) -- ( 42.00, 78.46);

\draw[color=drawColor,line cap=round,line join=round,] ( 48.00,101.69) -- ( 42.00,101.69);

\draw[color=drawColor,line cap=round,line join=round,] ( 48.00,124.92) -- ( 42.00,124.92);

\draw[color=drawColor,line cap=round,line join=round,] ( 48.00,148.16) -- ( 42.00,148.16);

\draw[color=drawColor,line cap=round,line join=round,] ( 48.00,171.39) -- ( 42.00,171.39);

\node[rotate= 90.00,color=drawColor,anchor=base,inner sep=0pt, outer sep=0pt, scale=  1.00] at ( 36.00, 55.23) {0};

\node[rotate= 90.00,color=drawColor,anchor=base,inner sep=0pt, outer sep=0pt, scale=  1.00] at ( 36.00, 78.46) {2};

\node[rotate= 90.00,color=drawColor,anchor=base,inner sep=0pt, outer sep=0pt, scale=  1.00] at ( 36.00,101.69) {4};

\node[rotate= 90.00,color=drawColor,anchor=base,inner sep=0pt, outer sep=0pt, scale=  1.00] at ( 36.00,124.92) {6};

\node[rotate= 90.00,color=drawColor,anchor=base,inner sep=0pt, outer sep=0pt, scale=  1.00] at ( 36.00,148.16) {8};

\node[rotate= 90.00,color=drawColor,anchor=base,inner sep=0pt, outer sep=0pt, scale=  1.00] at ( 36.00,171.39) {10};
\end{scope}
\begin{scope}

\definecolor[named]{drawColor}{rgb}{0.00,0.00,0.00}

\draw[color=drawColor,line cap=round,line join=round,] (103.97, 50.40) -- (103.97,180.67);
\end{scope}

%% file: phylogenetic_example1.histogram2.tex
\definecolor[named]{drawColor}{rgb}{0.00,0.00,0.00}
\definecolor[named]{fillColor}{rgb}{1.00,1.00,1.00}
\fill[color=fillColor,] (0,0) rectangle (180.67,180.67);
\begin{scope}

\definecolor[named]{drawColor}{rgb}{0.00,0.00,0.00}

\node[color=drawColor,anchor=base,inner sep=0pt, outer sep=0pt, scale=  1.00] at (108.34,  2.40) {$\leftarrow|e_1|$\quad\quad\quad$|e_3|\rightarrow$};

\node[rotate= 90.00,color=drawColor,anchor=base,inner sep=0pt, outer sep=0pt, scale=  1.00] at ( 12.00,115.54) {Density estimate};
\end{scope}
\begin{scope}

\definecolor[named]{drawColor}{rgb}{0.00,0.00,0.00}

\draw[color=drawColor,line cap=round,line join=round,] ( 52.47, 55.23) rectangle ( 54.75, 55.74);

\draw[color=drawColor,line cap=round,line join=round,] ( 54.75, 55.23) rectangle ( 57.03, 57.39);

\draw[color=drawColor,line cap=round,line join=round,] ( 57.03, 55.23) rectangle ( 59.31, 57.78);

\draw[color=drawColor,line cap=round,line join=round,] ( 59.31, 55.23) rectangle ( 61.59, 64.80);

\draw[color=drawColor,line cap=round,line join=round,] ( 61.59, 55.23) rectangle ( 63.87, 66.46);

\draw[color=drawColor,line cap=round,line join=round,] ( 63.87, 55.23) rectangle ( 66.15, 70.80);

\draw[color=drawColor,line cap=round,line join=round,] ( 66.15, 55.23) rectangle ( 68.43, 75.14);

\draw[color=drawColor,line cap=round,line join=round,] ( 68.43, 55.23) rectangle ( 70.71, 76.92);

\draw[color=drawColor,line cap=round,line join=round,] ( 70.71, 55.23) rectangle ( 72.99, 79.86);

\draw[color=drawColor,line cap=round,line join=round,] ( 72.99, 55.23) rectangle ( 75.27, 93.39);

\draw[color=drawColor,line cap=round,line join=round,] ( 75.27, 55.23) rectangle ( 77.55, 96.97);

\draw[color=drawColor,line cap=round,line join=round,] ( 77.55, 55.23) rectangle ( 79.83,111.01);

\draw[color=drawColor,line cap=round,line join=round,] ( 79.83, 55.23) rectangle ( 82.11,130.15);

\draw[color=drawColor,line cap=round,line join=round,] ( 82.11, 55.23) rectangle ( 84.39,139.09);

\draw[color=drawColor,line cap=round,line join=round,] ( 84.39, 55.23) rectangle ( 86.67,150.45);

\draw[color=drawColor,line cap=round,line join=round,] ( 86.67, 55.23) rectangle ( 88.95,163.47);

\draw[color=drawColor,line cap=round,line join=round,] ( 88.95, 55.23) rectangle ( 91.24,170.87);

\draw[color=drawColor,line cap=round,line join=round,] ( 91.24, 55.23) rectangle ( 93.52,169.98);

\draw[color=drawColor,line cap=round,line join=round,] ( 93.52, 55.23) rectangle ( 95.80,175.85);

\draw[color=drawColor,line cap=round,line join=round,] ( 95.80, 55.23) rectangle ( 98.08,163.47);

\draw[color=drawColor,line cap=round,line join=round,] ( 98.08, 55.23) rectangle (100.36,143.05);

\draw[color=drawColor,line cap=round,line join=round,] (100.36, 55.23) rectangle (102.64,129.77);

\draw[color=drawColor,line cap=round,line join=round,] (102.64, 55.23) rectangle (104.92,105.77);

\draw[color=drawColor,line cap=round,line join=round,] (104.92, 55.23) rectangle (107.20, 90.07);

\draw[color=drawColor,line cap=round,line join=round,] (107.20, 55.23) rectangle (109.48, 70.41);

\draw[color=drawColor,line cap=round,line join=round,] (109.48, 55.23) rectangle (111.76, 60.59);

\draw[color=drawColor,line cap=round,line join=round,] (111.76, 55.23) rectangle (114.04, 62.25);

\draw[color=drawColor,line cap=round,line join=round,] (114.04, 55.23) rectangle (116.32, 62.88);

\draw[color=drawColor,line cap=round,line join=round,] (116.32, 55.23) rectangle (118.60, 63.78);

\draw[color=drawColor,line cap=round,line join=round,] (118.60, 55.23) rectangle (120.88, 64.03);

\draw[color=drawColor,line cap=round,line join=round,] (120.88, 55.23) rectangle (123.16, 65.56);

\draw[color=drawColor,line cap=round,line join=round,] (123.16, 55.23) rectangle (125.44, 62.88);

\draw[color=drawColor,line cap=round,line join=round,] (125.44, 55.23) rectangle (127.72, 60.71);

\draw[color=drawColor,line cap=round,line join=round,] (127.72, 55.23) rectangle (130.00, 60.08);

\draw[color=drawColor,line cap=round,line join=round,] (130.00, 55.23) rectangle (132.28, 58.67);

\draw[color=drawColor,line cap=round,line join=round,] (132.28, 55.23) rectangle (134.56, 58.80);

\draw[color=drawColor,line cap=round,line join=round,] (134.56, 55.23) rectangle (136.84, 57.65);

\draw[color=drawColor,line cap=round,line join=round,] (136.84, 55.23) rectangle (139.12, 58.29);

\draw[color=drawColor,line cap=round,line join=round,] (139.12, 55.23) rectangle (141.40, 57.78);

\draw[color=drawColor,line cap=round,line join=round,] (141.40, 55.23) rectangle (143.68, 55.61);

\draw[color=drawColor,line cap=round,line join=round,] (143.68, 55.23) rectangle (145.96, 55.74);

\draw[color=drawColor,line cap=round,line join=round,] (145.96, 55.23) rectangle (148.24, 55.48);

\draw[color=drawColor,line cap=round,line join=round,] (148.24, 55.23) rectangle (150.52, 55.23);

\draw[color=drawColor,line cap=round,line join=round,] (150.52, 55.23) rectangle (152.80, 55.23);

\draw[color=drawColor,line cap=round,line join=round,] (152.80, 55.23) rectangle (155.08, 55.23);

\draw[color=drawColor,line cap=round,line join=round,] (155.08, 55.23) rectangle (157.36, 55.61);

\draw[color=drawColor,line cap=round,line join=round,] (157.36, 55.23) rectangle (159.64, 55.48);

\draw[color=drawColor,line cap=round,line join=round,] (159.64, 55.23) rectangle (161.93, 55.23);

\draw[color=drawColor,line cap=round,line join=round,] (161.93, 55.23) rectangle (164.21, 55.23);

\draw[color=drawColor,line cap=round,line join=round,] ( 43.94, 55.23) --
	( 44.19, 55.23) --
	( 44.43, 55.23) --
	( 44.67, 55.23) --
	( 44.91, 55.23) --
	( 45.15, 55.23) --
	( 45.39, 55.23) --
	( 45.63, 55.24) --
	( 45.87, 55.24) --
	( 46.11, 55.24) --
	( 46.35, 55.24) --
	( 46.59, 55.25) --
	( 46.83, 55.25) --
	( 47.07, 55.26) --
	( 47.31, 55.27) --
	( 47.55, 55.27) --
	( 47.80, 55.28) --
	( 48.04, 55.29) --
	( 48.28, 55.31) --
	( 48.52, 55.32) --
	( 48.76, 55.34) --
	( 49.00, 55.35) --
	( 49.24, 55.37) --
	( 49.48, 55.40) --
	( 49.72, 55.42) --
	( 49.96, 55.45) --
	( 50.20, 55.48) --
	( 50.44, 55.51) --
	( 50.68, 55.55) --
	( 50.92, 55.59) --
	( 51.17, 55.64) --
	( 51.41, 55.69) --
	( 51.65, 55.74) --
	( 51.89, 55.80) --
	( 52.13, 55.87) --
	( 52.37, 55.94) --
	( 52.61, 56.02) --
	( 52.85, 56.10) --
	( 53.09, 56.19) --
	( 53.33, 56.28) --
	( 53.57, 56.39) --
	( 53.81, 56.50) --
	( 54.05, 56.62) --
	( 54.29, 56.75) --
	( 54.54, 56.89) --
	( 54.78, 57.03) --
	( 55.02, 57.19) --
	( 55.26, 57.35) --
	( 55.50, 57.53) --
	( 55.74, 57.72) --
	( 55.98, 57.92) --
	( 56.22, 58.14) --
	( 56.46, 58.36) --
	( 56.70, 58.59) --
	( 56.94, 58.84) --
	( 57.18, 59.10) --
	( 57.42, 59.37) --
	( 57.66, 59.65) --
	( 57.91, 59.95) --
	( 58.15, 60.25) --
	( 58.39, 60.57) --
	( 58.63, 60.89) --
	( 58.87, 61.23) --
	( 59.11, 61.57) --
	( 59.35, 61.92) --
	( 59.59, 62.28) --
	( 59.83, 62.64) --
	( 60.07, 63.01) --
	( 60.31, 63.39) --
	( 60.55, 63.77) --
	( 60.79, 64.15) --
	( 61.03, 64.53) --
	( 61.27, 64.92) --
	( 61.52, 65.31) --
	( 61.76, 65.69) --
	( 62.00, 66.08) --
	( 62.24, 66.47) --
	( 62.48, 66.85) --
	( 62.72, 67.23) --
	( 62.96, 67.61) --
	( 63.20, 67.99) --
	( 63.44, 68.37) --
	( 63.68, 68.74) --
	( 63.92, 69.11) --
	( 64.16, 69.48) --
	( 64.40, 69.85) --
	( 64.64, 70.21) --
	( 64.89, 70.58) --
	( 65.13, 70.94) --
	( 65.37, 71.31) --
	( 65.61, 71.67) --
	( 65.85, 72.04) --
	( 66.09, 72.41) --
	( 66.33, 72.78) --
	( 66.57, 73.15) --
	( 66.81, 73.53) --
	( 67.05, 73.92) --
	( 67.29, 74.31) --
	( 67.53, 74.72) --
	( 67.77, 75.13) --
	( 68.01, 75.55) --
	( 68.26, 75.98) --
	( 68.50, 76.43) --
	( 68.74, 76.89) --
	( 68.98, 77.37) --
	( 69.22, 77.86) --
	( 69.46, 78.36) --
	( 69.70, 78.89) --
	( 69.94, 79.43) --
	( 70.18, 79.99) --
	( 70.42, 80.58) --
	( 70.66, 81.17) --
	( 70.90, 81.79) --
	( 71.14, 82.43) --
	( 71.38, 83.09) --
	( 71.63, 83.76) --
	( 71.87, 84.46) --
	( 72.11, 85.17) --
	( 72.35, 85.90) --
	( 72.59, 86.65) --
	( 72.83, 87.42) --
	( 73.07, 88.21) --
	( 73.31, 89.01) --
	( 73.55, 89.84) --
	( 73.79, 90.68) --
	( 74.03, 91.54) --
	( 74.27, 92.42) --
	( 74.51, 93.32) --
	( 74.75, 94.25) --
	( 74.99, 95.19) --
	( 75.24, 96.16) --
	( 75.48, 97.16) --
	( 75.72, 98.18) --
	( 75.96, 99.23) --
	( 76.20,100.30) --
	( 76.44,101.40) --
	( 76.68,102.53) --
	( 76.92,103.69) --
	( 77.16,104.88) --
	( 77.40,106.10) --
	( 77.64,107.34) --
	( 77.88,108.61) --
	( 78.12,109.90) --
	( 78.36,111.22) --
	( 78.61,112.56) --
	( 78.85,113.92) --
	( 79.09,115.29) --
	( 79.33,116.68) --
	( 79.57,118.08) --
	( 79.81,119.48) --
	( 80.05,120.89) --
	( 80.29,122.30) --
	( 80.53,123.71) --
	( 80.77,125.12) --
	( 81.01,126.52) --
	( 81.25,127.91) --
	( 81.49,129.29) --
	( 81.73,130.65) --
	( 81.98,132.01) --
	( 82.22,133.35) --
	( 82.46,134.68) --
	( 82.70,135.99) --
	( 82.94,137.28) --
	( 83.18,138.55) --
	( 83.42,139.81) --
	( 83.66,141.06) --
	( 83.90,142.29) --
	( 84.14,143.50) --
	( 84.38,144.69) --
	( 84.62,145.87) --
	( 84.86,147.04) --
	( 85.10,148.18) --
	( 85.35,149.31) --
	( 85.59,150.43) --
	( 85.83,151.52) --
	( 86.07,152.59) --
	( 86.31,153.64) --
	( 86.55,154.68) --
	( 86.79,155.69) --
	( 87.03,156.67) --
	( 87.27,157.62) --
	( 87.51,158.55) --
	( 87.75,159.45) --
	( 87.99,160.31) --
	( 88.23,161.15) --
	( 88.47,161.94) --
	( 88.71,162.70) --
	( 88.96,163.43) --
	( 89.20,164.12) --
	( 89.44,164.77) --
	( 89.68,165.38) --
	( 89.92,165.94) --
	( 90.16,166.46) --
	( 90.40,166.94) --
	( 90.64,167.37) --
	( 90.88,167.77) --
	( 91.12,168.11) --
	( 91.36,168.41) --
	( 91.60,168.65) --
	( 91.84,168.85) --
	( 92.08,169.00) --
	( 92.33,169.10) --
	( 92.57,169.14) --
	( 92.81,169.11) --
	( 93.05,169.03) --
	( 93.29,168.89) --
	( 93.53,168.69) --
	( 93.77,168.42) --
	( 94.01,168.08) --
	( 94.25,167.67) --
	( 94.49,167.18) --
	( 94.73,166.62) --
	( 94.97,166.00) --
	( 95.21,165.29) --
	( 95.45,164.52) --
	( 95.70,163.65) --
	( 95.94,162.72) --
	( 96.18,161.71) --
	( 96.42,160.63) --
	( 96.66,159.48) --
	( 96.90,158.26) --
	( 97.14,156.97) --
	( 97.38,155.62) --
	( 97.62,154.21) --
	( 97.86,152.74) --
	( 98.10,151.22) --
	( 98.34,149.66) --
	( 98.58,148.04) --
	( 98.82,146.38) --
	( 99.07,144.68) --
	( 99.31,142.96) --
	( 99.55,141.20) --
	( 99.79,139.41) --
	(100.03,137.60) --
	(100.27,135.76) --
	(100.51,133.91) --
	(100.75,132.05) --
	(100.99,130.17) --
	(101.23,128.28) --
	(101.47,126.38) --
	(101.71,124.48) --
	(101.95,122.57) --
	(102.19,120.66) --
	(102.43,118.75) --
	(102.68,116.84) --
	(102.92,114.93) --
	(103.16,113.02) --
	(103.40,111.13) --
	(103.64,109.23) --
	(103.88,107.35) --
	(104.12,105.48) --
	(104.36,103.62) --
	(104.60,101.78) --
	(104.84, 99.96) --
	(105.08, 98.15) --
	(105.32, 96.36) --
	(105.56, 94.59) --
	(105.80, 92.86) --
	(106.05, 91.15) --
	(106.29, 89.47) --
	(106.53, 87.83) --
	(106.77, 86.22) --
	(107.01, 84.65) --
	(107.25, 83.13) --
	(107.49, 81.65) --
	(107.73, 80.22) --
	(107.97, 78.84) --
	(108.21, 77.51) --
	(108.45, 76.24) --
	(108.69, 75.04) --
	(108.93, 73.89) --
	(109.17, 72.81) --
	(109.42, 71.78) --
	(109.66, 70.82) --
	(109.90, 69.92) --
	(110.14, 69.10) --
	(110.38, 68.34) --
	(110.62, 67.63) --
	(110.86, 66.99) --
	(111.10, 66.41) --
	(111.34, 65.89) --
	(111.58, 65.43) --
	(111.82, 65.02) --
	(112.06, 64.66) --
	(112.30, 64.35) --
	(112.54, 64.08) --
	(112.79, 63.85) --
	(113.03, 63.66) --
	(113.27, 63.51) --
	(113.51, 63.39) --
	(113.75, 63.29) --
	(113.99, 63.22) --
	(114.23, 63.17) --
	(114.47, 63.15) --
	(114.71, 63.14) --
	(114.95, 63.14) --
	(115.19, 63.16) --
	(115.43, 63.19) --
	(115.67, 63.22) --
	(115.91, 63.27) --
	(116.15, 63.32) --
	(116.40, 63.37) --
	(116.64, 63.43) --
	(116.88, 63.49) --
	(117.12, 63.55) --
	(117.36, 63.61) --
	(117.60, 63.67) --
	(117.84, 63.73) --
	(118.08, 63.79) --
	(118.32, 63.84) --
	(118.56, 63.89) --
	(118.80, 63.93) --
	(119.04, 63.97) --
	(119.28, 64.00) --
	(119.52, 64.03) --
	(119.77, 64.05) --
	(120.01, 64.05) --
	(120.25, 64.05) --
	(120.49, 64.05) --
	(120.73, 64.03) --
	(120.97, 64.00) --
	(121.21, 63.96) --
	(121.45, 63.91) --
	(121.69, 63.85) --
	(121.93, 63.78) --
	(122.17, 63.70) --
	(122.41, 63.61) --
	(122.65, 63.51) --
	(122.89, 63.40) --
	(123.14, 63.29) --
	(123.38, 63.16) --
	(123.62, 63.03) --
	(123.86, 62.90) --
	(124.10, 62.76) --
	(124.34, 62.62) --
	(124.58, 62.48) --
	(124.82, 62.33) --
	(125.06, 62.18) --
	(125.30, 62.04) --
	(125.54, 61.89) --
	(125.78, 61.75) --
	(126.02, 61.60) --
	(126.26, 61.46) --
	(126.51, 61.33) --
	(126.75, 61.19) --
	(126.99, 61.07) --
	(127.23, 60.94) --
	(127.47, 60.82) --
	(127.71, 60.70) --
	(127.95, 60.59) --
	(128.19, 60.48) --
	(128.43, 60.37) --
	(128.67, 60.27) --
	(128.91, 60.17) --
	(129.15, 60.07) --
	(129.39, 59.97) --
	(129.63, 59.88) --
	(129.87, 59.79) --
	(130.12, 59.71) --
	(130.36, 59.62) --
	(130.60, 59.54) --
	(130.84, 59.46) --
	(131.08, 59.38) --
	(131.32, 59.30) --
	(131.56, 59.22) --
	(131.80, 59.15) --
	(132.04, 59.08) --
	(132.28, 59.01) --
	(132.52, 58.94) --
	(132.76, 58.87) --
	(133.00, 58.81) --
	(133.24, 58.75) --
	(133.49, 58.69) --
	(133.73, 58.63) --
	(133.97, 58.57) --
	(134.21, 58.52) --
	(134.45, 58.47) --
	(134.69, 58.42) --
	(134.93, 58.37) --
	(135.17, 58.32) --
	(135.41, 58.27) --
	(135.65, 58.22) --
	(135.89, 58.18) --
	(136.13, 58.13) --
	(136.37, 58.08) --
	(136.61, 58.04) --
	(136.86, 57.99) --
	(137.10, 57.94) --
	(137.34, 57.89) --
	(137.58, 57.84) --
	(137.82, 57.78) --
	(138.06, 57.73) --
	(138.30, 57.67) --
	(138.54, 57.61) --
	(138.78, 57.55) --
	(139.02, 57.48) --
	(139.26, 57.41) --
	(139.50, 57.34) --
	(139.74, 57.27) --
	(139.98, 57.20) --
	(140.23, 57.12) --
	(140.47, 57.05) --
	(140.71, 56.97) --
	(140.95, 56.89) --
	(141.19, 56.81) --
	(141.43, 56.74) --
	(141.67, 56.66) --
	(141.91, 56.58) --
	(142.15, 56.51) --
	(142.39, 56.44) --
	(142.63, 56.37) --
	(142.87, 56.30) --
	(143.11, 56.23) --
	(143.35, 56.17) --
	(143.59, 56.11) --
	(143.84, 56.05) --
	(144.08, 56.00) --
	(144.32, 55.94) --
	(144.56, 55.89) --
	(144.80, 55.85) --
	(145.04, 55.81) --
	(145.28, 55.76) --
	(145.52, 55.73) --
	(145.76, 55.69) --
	(146.00, 55.66) --
	(146.24, 55.63) --
	(146.48, 55.60) --
	(146.72, 55.57) --
	(146.96, 55.54) --
	(147.21, 55.52) --
	(147.45, 55.50) --
	(147.69, 55.48) --
	(147.93, 55.46) --
	(148.17, 55.44) --
	(148.41, 55.43) --
	(148.65, 55.41) --
	(148.89, 55.40) --
	(149.13, 55.39) --
	(149.37, 55.38) --
	(149.61, 55.37) --
	(149.85, 55.36) --
	(150.09, 55.36) --
	(150.33, 55.35) --
	(150.58, 55.35) --
	(150.82, 55.34) --
	(151.06, 55.34) --
	(151.30, 55.34) --
	(151.54, 55.34) --
	(151.78, 55.34) --
	(152.02, 55.34) --
	(152.26, 55.35) --
	(152.50, 55.35) --
	(152.74, 55.35) --
	(152.98, 55.36) --
	(153.22, 55.36) --
	(153.46, 55.37) --
	(153.70, 55.37) --
	(153.95, 55.38) --
	(154.19, 55.38) --
	(154.43, 55.39) --
	(154.67, 55.39) --
	(154.91, 55.39) --
	(155.15, 55.40) --
	(155.39, 55.40) --
	(155.63, 55.40) --
	(155.87, 55.40) --
	(156.11, 55.40) --
	(156.35, 55.40) --
	(156.59, 55.40) --
	(156.83, 55.40) --
	(157.07, 55.39) --
	(157.31, 55.39) --
	(157.56, 55.39) --
	(157.80, 55.38) --
	(158.04, 55.37) --
	(158.28, 55.37) --
	(158.52, 55.36) --
	(158.76, 55.35) --
	(159.00, 55.35) --
	(159.24, 55.34) --
	(159.48, 55.33) --
	(159.72, 55.32) --
	(159.96, 55.32) --
	(160.20, 55.31) --
	(160.44, 55.30) --
	(160.68, 55.29) --
	(160.93, 55.29) --
	(161.17, 55.28) --
	(161.41, 55.28) --
	(161.65, 55.27) --
	(161.89, 55.26) --
	(162.13, 55.26) --
	(162.37, 55.26) --
	(162.61, 55.25) --
	(162.85, 55.25) --
	(163.09, 55.24) --
	(163.33, 55.24) --
	(163.57, 55.24) --
	(163.81, 55.24) --
	(164.05, 55.24) --
	(164.30, 55.23) --
	(164.54, 55.23) --
	(164.78, 55.23) --
	(165.02, 55.23) --
	(165.26, 55.23) --
	(165.50, 55.23) --
	(165.74, 55.23) --
	(165.98, 55.23) --
	(166.22, 55.23) --
	(166.46, 55.23) --
	(166.70, 55.23) --
	(166.94, 55.23);
\end{scope}
\begin{scope}

\definecolor[named]{drawColor}{rgb}{0.00,0.00,0.00}

\draw[color=drawColor,line cap=round,line join=round,] ( 52.47, 50.40) -- (164.21, 50.40);

\draw[color=drawColor,line cap=round,line join=round,] ( 52.47, 50.40) -- ( 52.47, 44.40);

\draw[color=drawColor,line cap=round,line join=round,] ( 80.40, 50.40) -- ( 80.40, 44.40);

\draw[color=drawColor,line cap=round,line join=round,] (108.34, 50.40) -- (108.34, 44.40);

\draw[color=drawColor,line cap=round,line join=round,] (136.27, 50.40) -- (136.27, 44.40);

\draw[color=drawColor,line cap=round,line join=round,] (164.21, 50.40) -- (164.21, 44.40);

\node[color=drawColor,anchor=base,inner sep=0pt, outer sep=0pt, scale=  1.00] at ( 52.47, 26.40) {0.2};

\node[color=drawColor,anchor=base,inner sep=0pt, outer sep=0pt, scale=  1.00] at ( 80.40, 26.40) {0.1};

\node[color=drawColor,anchor=base,inner sep=0pt, outer sep=0pt, scale=  1.00] at (108.34, 26.40) {0};

\node[color=drawColor,anchor=base,inner sep=0pt, outer sep=0pt, scale=  1.00] at (136.27, 26.40) {0.1};

\node[color=drawColor,anchor=base,inner sep=0pt, outer sep=0pt, scale=  1.00] at (164.21, 26.40) {0.2};

\draw[color=drawColor,line cap=round,line join=round,] ( 48.00, 55.23) -- ( 48.00,170.38);

\draw[color=drawColor,line cap=round,line join=round,] ( 48.00, 55.23) -- ( 42.00, 55.23);

\draw[color=drawColor,line cap=round,line join=round,] ( 48.00, 78.26) -- ( 42.00, 78.26);

\draw[color=drawColor,line cap=round,line join=round,] ( 48.00,101.29) -- ( 42.00,101.29);

\draw[color=drawColor,line cap=round,line join=round,] ( 48.00,124.32) -- ( 42.00,124.32);

\draw[color=drawColor,line cap=round,line join=round,] ( 48.00,147.35) -- ( 42.00,147.35);

\draw[color=drawColor,line cap=round,line join=round,] ( 48.00,170.38) -- ( 42.00,170.38);

\node[rotate= 90.00,color=drawColor,anchor=base,inner sep=0pt, outer sep=0pt, scale=  1.00] at ( 36.00, 55.23) {0};

\node[rotate= 90.00,color=drawColor,anchor=base,inner sep=0pt, outer sep=0pt, scale=  1.00] at ( 36.00, 78.26) {2};

\node[rotate= 90.00,color=drawColor,anchor=base,inner sep=0pt, outer sep=0pt, scale=  1.00] at ( 36.00,101.29) {4};

\node[rotate= 90.00,color=drawColor,anchor=base,inner sep=0pt, outer sep=0pt, scale=  1.00] at ( 36.00,124.32) {6};

\node[rotate= 90.00,color=drawColor,anchor=base,inner sep=0pt, outer sep=0pt, scale=  1.00] at ( 36.00,147.35) {8};

\node[rotate= 90.00,color=drawColor,anchor=base,inner sep=0pt, outer sep=0pt, scale=  1.00] at ( 36.00,170.38) {10};
\end{scope}
\begin{scope}

\definecolor[named]{drawColor}{rgb}{0.00,0.00,0.00}

\draw[color=drawColor,line cap=round,line join=round,] (103.97, 50.40) -- (103.97,180.67);
\end{scope}

%% file: phylogenetic_example2.mean.tex
\definecolor[named]{drawColor}{rgb}{0.00,0.00,0.00}
\definecolor[named]{fillColor}{rgb}{1.00,1.00,1.00}
\fill[color=fillColor,] (0,0) rectangle (216.81,216.81);
\begin{scope}

\definecolor[named]{drawColor}{rgb}{0.00,0.00,0.00}

\draw[color=drawColor,line cap=round,line join=round,] (  8.03,163.99) -- ( 24.04, 91.15);

\draw[color=drawColor,line cap=round,line join=round,] ( 24.04, 91.15) -- ( 27.41, 35.21);

\draw[color=drawColor,line cap=round,line join=round,] ( 27.41, 35.21) -- ( 27.43, 16.39);

\draw[color=drawColor,line cap=round,line join=round,] ( 27.43, 16.39) -- ( 66.17,  8.03);

\draw[color=drawColor,line cap=round,line join=round,] ( 27.43, 16.39) -- ( 67.10, 24.76);

\draw[color=drawColor,line cap=round,line join=round,] ( 27.41, 35.21) -- ( 27.51, 54.04);

\draw[color=drawColor,line cap=round,line join=round,] ( 27.51, 54.04) -- ( 73.53, 41.49);

\draw[color=drawColor,line cap=round,line join=round,] ( 27.51, 54.04) -- ( 42.38, 66.58);

\draw[color=drawColor,line cap=round,line join=round,] ( 42.38, 66.58) -- ( 67.83, 58.22);

\draw[color=drawColor,line cap=round,line join=round,] ( 42.38, 66.58) -- ( 58.69, 74.95);

\draw[color=drawColor,line cap=round,line join=round,] ( 24.04, 91.15) -- ( 60.35,147.09);

\draw[color=drawColor,line cap=round,line join=round,] ( 60.35,147.09) -- ( 60.46,127.23);

\draw[color=drawColor,line cap=round,line join=round,] ( 60.46,127.23) -- ( 64.04,112.59);

\draw[color=drawColor,line cap=round,line join=round,] ( 64.04,112.59) -- ( 66.15,100.04);

\draw[color=drawColor,line cap=round,line join=round,] ( 66.15,100.04) -- ( 72.97, 91.68);

\draw[color=drawColor,line cap=round,line join=round,] ( 66.15,100.04) -- ( 68.76,108.40);

\draw[color=drawColor,line cap=round,line join=round,] ( 64.04,112.59) -- ( 70.43,125.13);

\draw[color=drawColor,line cap=round,line join=round,] ( 60.46,127.23) -- ( 67.64,141.86);

\draw[color=drawColor,line cap=round,line join=round,] ( 60.35,147.09) -- (102.17,166.96);

\draw[color=drawColor,line cap=round,line join=round,] (102.17,166.96) -- (172.75,158.59);

\draw[color=drawColor,line cap=round,line join=round,] (102.17,166.96) -- (108.60,175.32);

\draw[color=drawColor,line cap=round,line join=round,] (  8.03,163.99) -- ( 37.27,192.05);

\draw[color=drawColor,line cap=round,line join=round,] (  8.03,163.99) -- ( 31.31,208.78);

\node[color=drawColor,anchor=base west,inner sep=0pt, outer sep=0pt, scale=  1.00] at ( 66.17,  5.56) {\itshape Squirrel};

\node[color=drawColor,anchor=base west,inner sep=0pt, outer sep=0pt, scale=  1.00] at ( 67.10, 21.32) {\itshape Guinea pig};

\node[color=drawColor,anchor=base west,inner sep=0pt, outer sep=0pt, scale=  1.00] at ( 73.53, 39.02) {\itshape Kangaroo rat};

\node[color=drawColor,anchor=base west,inner sep=0pt, outer sep=0pt, scale=  1.00] at ( 67.83, 55.01) {\itshape Mouse};

\node[color=drawColor,anchor=base west,inner sep=0pt, outer sep=0pt, scale=  1.00] at ( 58.69, 72.71) {\itshape Rat};

\node[color=drawColor,anchor=base west,inner sep=0pt, outer sep=0pt, scale=  1.00] at ( 72.97, 89.20) {\itshape Human};

\node[color=drawColor,anchor=base west,inner sep=0pt, outer sep=0pt, scale=  1.00] at ( 68.76,105.93) {\itshape Chimpanzee};

\node[color=drawColor,anchor=base west,inner sep=0pt, outer sep=0pt, scale=  1.00] at ( 70.43,122.66) {\itshape Gorilla};

\node[color=drawColor,anchor=base west,inner sep=0pt, outer sep=0pt, scale=  1.00] at ( 67.64,139.39) {\itshape Orangutan};

\node[color=drawColor,anchor=base west,inner sep=0pt, outer sep=0pt, scale=  1.00] at (172.75,155.15) {\itshape Marmoset};

\node[color=drawColor,anchor=base west,inner sep=0pt, outer sep=0pt, scale=  1.00] at (108.60,172.85) {\itshape Baboon};

\node[color=drawColor,anchor=base west,inner sep=0pt, outer sep=0pt, scale=  1.00] at ( 37.27,189.58) {\itshape Rabbit};

\node[color=drawColor,anchor=base west,inner sep=0pt, outer sep=0pt, scale=  1.00] at ( 31.31,205.34) {\itshape Pika};

\node[color=drawColor,anchor=base,inner sep=0pt, outer sep=0pt, scale=  1.00] at (  1.52, 55.01) {0.031};

\node[color=drawColor,anchor=base,inner sep=0pt, outer sep=0pt, scale=  1.00] at (-10.43,121.93) {0.148};
\end{scope}

%% file: phylogenetic_example2.consensus.tex
\definecolor[named]{drawColor}{rgb}{0.00,0.00,0.00}
\definecolor[named]{fillColor}{rgb}{1.00,1.00,1.00}
\fill[color=fillColor,] (0,0) rectangle (216.81,216.81);
\begin{scope}

\definecolor[named]{drawColor}{rgb}{0.00,0.00,0.00}

\draw[color=drawColor,line cap=round,line join=round,] (  8.03,162.43) -- ( 23.43, 86.45);

\draw[color=drawColor,line cap=round,line join=round,] ( 23.43, 86.45) -- ( 31.21, 35.21);

\draw[color=drawColor,line cap=round,line join=round,] ( 31.21, 35.21) -- ( 68.45,  8.03);

\draw[color=drawColor,line cap=round,line join=round,] ( 31.21, 35.21) -- ( 69.34, 24.76);

\draw[color=drawColor,line cap=round,line join=round,] ( 31.21, 35.21) -- ( 75.45, 41.49);

\draw[color=drawColor,line cap=round,line join=round,] ( 31.21, 35.21) -- ( 45.54, 66.58);

\draw[color=drawColor,line cap=round,line join=round,] ( 45.54, 66.58) -- ( 70.01, 58.22);

\draw[color=drawColor,line cap=round,line join=round,] ( 45.54, 66.58) -- ( 61.22, 74.95);

\draw[color=drawColor,line cap=round,line join=round,] ( 23.43, 86.45) -- ( 58.33,137.68);

\draw[color=drawColor,line cap=round,line join=round,] ( 58.33,137.68) -- ( 63.79,112.59);

\draw[color=drawColor,line cap=round,line join=round,] ( 63.79,112.59) -- ( 66.74,100.04);

\draw[color=drawColor,line cap=round,line join=round,] ( 66.74,100.04) -- ( 73.30, 91.68);

\draw[color=drawColor,line cap=round,line join=round,] ( 66.74,100.04) -- ( 69.25,108.40);

\draw[color=drawColor,line cap=round,line join=round,] ( 63.79,112.59) -- ( 69.93,125.13);

\draw[color=drawColor,line cap=round,line join=round,] ( 58.33,137.68) -- ( 64.71,162.77);

\draw[color=drawColor,line cap=round,line join=round,] ( 64.71,162.77) -- (104.91,150.23);

\draw[color=drawColor,line cap=round,line join=round,] (104.91,150.23) -- (172.75,141.86);

\draw[color=drawColor,line cap=round,line join=round,] (104.91,150.23) -- (111.09,158.59);

\draw[color=drawColor,line cap=round,line join=round,] ( 64.71,162.77) -- ( 71.62,175.32);

\draw[color=drawColor,line cap=round,line join=round,] (  8.03,162.43) -- ( 36.13,192.05);

\draw[color=drawColor,line cap=round,line join=round,] (  8.03,162.43) -- ( 30.41,208.78);

\node[color=drawColor,anchor=base west,inner sep=0pt, outer sep=0pt, scale=  1.00] at ( 68.45,  5.56) {\itshape Squirrel};

\node[color=drawColor,anchor=base west,inner sep=0pt, outer sep=0pt, scale=  1.00] at ( 69.34, 21.32) {\itshape Guinea pig};

\node[color=drawColor,anchor=base west,inner sep=0pt, outer sep=0pt, scale=  1.00] at ( 75.45, 39.02) {\itshape Kangaroo rat};

\node[color=drawColor,anchor=base west,inner sep=0pt, outer sep=0pt, scale=  1.00] at ( 70.01, 55.01) {\itshape Mouse};

\node[color=drawColor,anchor=base west,inner sep=0pt, outer sep=0pt, scale=  1.00] at ( 61.22, 72.71) {\itshape Rat};

\node[color=drawColor,anchor=base west,inner sep=0pt, outer sep=0pt, scale=  1.00] at ( 73.30, 89.20) {\itshape Human};

\node[color=drawColor,anchor=base west,inner sep=0pt, outer sep=0pt, scale=  1.00] at ( 69.25,105.93) {\itshape Chimpanzee};

\node[color=drawColor,anchor=base west,inner sep=0pt, outer sep=0pt, scale=  1.00] at ( 69.93,122.66) {\itshape Gorilla};

\node[color=drawColor,anchor=base west,inner sep=0pt, outer sep=0pt, scale=  1.00] at (172.75,138.42) {\itshape Marmoset};

\node[color=drawColor,anchor=base west,inner sep=0pt, outer sep=0pt, scale=  1.00] at (111.09,156.12) {\itshape Baboon};

\node[color=drawColor,anchor=base west,inner sep=0pt, outer sep=0pt, scale=  1.00] at ( 71.62,172.85) {\itshape Orangutan};

\node[color=drawColor,anchor=base west,inner sep=0pt, outer sep=0pt, scale=  1.00] at ( 36.13,189.58) {\itshape Rabbit};

\node[color=drawColor,anchor=base west,inner sep=0pt, outer sep=0pt, scale=  1.00] at ( 30.41,205.34) {\itshape Pika};

\node[color=drawColor,anchor=base,inner sep=0pt, outer sep=0pt, scale=  1.00] at (  1.77, 55.01) {0.074};

\node[color=drawColor,anchor=base,inner sep=0pt, outer sep=0pt, scale=  1.00] at (-10.75,121.93) {0.148};
\end{scope}

%% file: phylogenetic_example2.histogram1.tex
\definecolor[named]{drawColor}{rgb}{0.00,0.00,0.00}
\definecolor[named]{fillColor}{rgb}{1.00,1.00,1.00}
\fill[color=fillColor,] (0,0) rectangle (180.67,180.67);
\begin{scope}

\definecolor[named]{drawColor}{rgb}{0.00,0.00,0.00}

\node[color=drawColor,anchor=base,inner sep=0pt, outer sep=0pt, scale=  1.00] at (108.34,  2.40) {$\leftarrow|e_4|$\quad\quad\quad$|e_5|\rightarrow$};

\node[rotate= 90.00,color=drawColor,anchor=base,inner sep=0pt, outer sep=0pt, scale=  1.00] at ( 12.00,115.54) {Density estimate};
\end{scope}
\begin{scope}

\definecolor[named]{drawColor}{rgb}{0.00,0.00,0.00}

\draw[color=drawColor,line cap=round,line join=round,] ( 52.47, 55.23) rectangle ( 54.75, 58.12);

\draw[color=drawColor,line cap=round,line join=round,] ( 54.75, 55.23) rectangle ( 57.03, 60.30);

\draw[color=drawColor,line cap=round,line join=round,] ( 57.03, 55.23) rectangle ( 59.31, 63.34);

\draw[color=drawColor,line cap=round,line join=round,] ( 59.31, 55.23) rectangle ( 61.59, 63.15);

\draw[color=drawColor,line cap=round,line join=round,] ( 61.59, 55.23) rectangle ( 63.87, 65.19);

\draw[color=drawColor,line cap=round,line join=round,] ( 63.87, 55.23) rectangle ( 66.15, 71.98);

\draw[color=drawColor,line cap=round,line join=round,] ( 66.15, 55.23) rectangle ( 68.43, 75.26);

\draw[color=drawColor,line cap=round,line join=round,] ( 68.43, 55.23) rectangle ( 70.71, 78.72);

\draw[color=drawColor,line cap=round,line join=round,] ( 70.71, 55.23) rectangle ( 72.99, 84.56);

\draw[color=drawColor,line cap=round,line join=round,] ( 72.99, 55.23) rectangle ( 75.27, 90.35);

\draw[color=drawColor,line cap=round,line join=round,] ( 75.27, 55.23) rectangle ( 77.55, 98.52);

\draw[color=drawColor,line cap=round,line join=round,] ( 77.55, 55.23) rectangle ( 79.83,112.71);

\draw[color=drawColor,line cap=round,line join=round,] ( 79.83, 55.23) rectangle ( 82.11,123.20);

\draw[color=drawColor,line cap=round,line join=round,] ( 82.11, 55.23) rectangle ( 84.39,137.21);

\draw[color=drawColor,line cap=round,line join=round,] ( 84.39, 55.23) rectangle ( 86.67,143.95);

\draw[color=drawColor,line cap=round,line join=round,] ( 86.67, 55.23) rectangle ( 88.95,153.16);

\draw[color=drawColor,line cap=round,line join=round,] ( 88.95, 55.23) rectangle ( 91.24,167.92);

\draw[color=drawColor,line cap=round,line join=round,] ( 91.24, 55.23) rectangle ( 93.52,162.80);

\draw[color=drawColor,line cap=round,line join=round,] ( 93.52, 55.23) rectangle ( 95.80,175.85);

\draw[color=drawColor,line cap=round,line join=round,] ( 95.80, 55.23) rectangle ( 98.08,171.58);

\draw[color=drawColor,line cap=round,line join=round,] ( 98.08, 55.23) rectangle (100.36,160.66);

\draw[color=drawColor,line cap=round,line join=round,] (100.36, 55.23) rectangle (102.64,153.87);

\draw[color=drawColor,line cap=round,line join=round,] (102.64, 55.23) rectangle (104.92,149.31);

\draw[color=drawColor,line cap=round,line join=round,] (104.92, 55.23) rectangle (107.20,133.17);

\draw[color=drawColor,line cap=round,line join=round,] (107.20, 55.23) rectangle (109.48,106.40);

\draw[color=drawColor,line cap=round,line join=round,] (109.48, 55.23) rectangle (111.76, 95.29);

\draw[color=drawColor,line cap=round,line join=round,] (111.76, 55.23) rectangle (114.04, 96.62);

\draw[color=drawColor,line cap=round,line join=round,] (114.04, 55.23) rectangle (116.32, 96.53);

\draw[color=drawColor,line cap=round,line join=round,] (116.32, 55.23) rectangle (118.60, 95.34);

\draw[color=drawColor,line cap=round,line join=round,] (118.60, 55.23) rectangle (120.88, 93.20);

\draw[color=drawColor,line cap=round,line join=round,] (120.88, 55.23) rectangle (123.16, 85.42);

\draw[color=drawColor,line cap=round,line join=round,] (123.16, 55.23) rectangle (125.44, 81.52);

\draw[color=drawColor,line cap=round,line join=round,] (125.44, 55.23) rectangle (127.72, 77.54);

\draw[color=drawColor,line cap=round,line join=round,] (127.72, 55.23) rectangle (130.00, 73.22);

\draw[color=drawColor,line cap=round,line join=round,] (130.00, 55.23) rectangle (132.28, 71.70);

\draw[color=drawColor,line cap=round,line join=round,] (132.28, 55.23) rectangle (134.56, 66.62);

\draw[color=drawColor,line cap=round,line join=round,] (134.56, 55.23) rectangle (136.84, 64.53);

\draw[color=drawColor,line cap=round,line join=round,] (136.84, 55.23) rectangle (139.12, 61.16);

\draw[color=drawColor,line cap=round,line join=round,] (139.12, 55.23) rectangle (141.40, 60.26);

\draw[color=drawColor,line cap=round,line join=round,] (141.40, 55.23) rectangle (143.68, 59.78);

\draw[color=drawColor,line cap=round,line join=round,] (143.68, 55.23) rectangle (145.96, 57.60);

\draw[color=drawColor,line cap=round,line join=round,] (145.96, 55.23) rectangle (148.24, 56.74);

\draw[color=drawColor,line cap=round,line join=round,] (148.24, 55.23) rectangle (150.52, 56.22);

\draw[color=drawColor,line cap=round,line join=round,] (150.52, 55.23) rectangle (152.80, 56.70);

\draw[color=drawColor,line cap=round,line join=round,] (152.80, 55.23) rectangle (155.08, 55.84);

\draw[color=drawColor,line cap=round,line join=round,] (155.08, 55.23) rectangle (157.36, 55.70);

\draw[color=drawColor,line cap=round,line join=round,] (157.36, 55.23) rectangle (159.64, 55.46);

\draw[color=drawColor,line cap=round,line join=round,] (159.64, 55.23) rectangle (161.93, 55.37);

\draw[color=drawColor,line cap=round,line join=round,] (161.93, 55.23) rectangle (164.21, 55.23);

\draw[color=drawColor,line cap=round,line join=round,] ( 42.74, 55.23) --
	( 42.99, 55.23) --
	( 43.24, 55.23) --
	( 43.49, 55.23) --
	( 43.74, 55.23) --
	( 43.99, 55.24) --
	( 44.24, 55.24) --
	( 44.49, 55.24) --
	( 44.74, 55.25) --
	( 44.99, 55.26) --
	( 45.24, 55.26) --
	( 45.49, 55.27) --
	( 45.75, 55.28) --
	( 46.00, 55.29) --
	( 46.25, 55.31) --
	( 46.50, 55.32) --
	( 46.75, 55.34) --
	( 47.00, 55.37) --
	( 47.25, 55.39) --
	( 47.50, 55.42) --
	( 47.75, 55.46) --
	( 48.00, 55.50) --
	( 48.25, 55.55) --
	( 48.50, 55.60) --
	( 48.75, 55.65) --
	( 49.00, 55.72) --
	( 49.25, 55.79) --
	( 49.50, 55.87) --
	( 49.75, 55.95) --
	( 50.00, 56.04) --
	( 50.25, 56.15) --
	( 50.50, 56.26) --
	( 50.75, 56.37) --
	( 51.00, 56.50) --
	( 51.25, 56.64) --
	( 51.50, 56.78) --
	( 51.75, 56.93) --
	( 52.00, 57.09) --
	( 52.25, 57.26) --
	( 52.50, 57.44) --
	( 52.75, 57.62) --
	( 53.00, 57.81) --
	( 53.25, 58.00) --
	( 53.50, 58.20) --
	( 53.75, 58.41) --
	( 54.00, 58.61) --
	( 54.25, 58.83) --
	( 54.50, 59.04) --
	( 54.75, 59.26) --
	( 55.00, 59.48) --
	( 55.25, 59.70) --
	( 55.50, 59.93) --
	( 55.75, 60.15) --
	( 56.00, 60.37) --
	( 56.25, 60.60) --
	( 56.50, 60.82) --
	( 56.75, 61.05) --
	( 57.01, 61.27) --
	( 57.26, 61.50) --
	( 57.51, 61.72) --
	( 57.76, 61.95) --
	( 58.01, 62.17) --
	( 58.26, 62.40) --
	( 58.51, 62.63) --
	( 58.76, 62.86) --
	( 59.01, 63.10) --
	( 59.26, 63.34) --
	( 59.51, 63.59) --
	( 59.76, 63.84) --
	( 60.01, 64.10) --
	( 60.26, 64.36) --
	( 60.51, 64.64) --
	( 60.76, 64.92) --
	( 61.01, 65.21) --
	( 61.26, 65.51) --
	( 61.51, 65.82) --
	( 61.76, 66.14) --
	( 62.01, 66.47) --
	( 62.26, 66.81) --
	( 62.51, 67.16) --
	( 62.76, 67.52) --
	( 63.01, 67.89) --
	( 63.26, 68.27) --
	( 63.51, 68.66) --
	( 63.76, 69.05) --
	( 64.01, 69.45) --
	( 64.26, 69.86) --
	( 64.51, 70.28) --
	( 64.76, 70.70) --
	( 65.01, 71.12) --
	( 65.26, 71.55) --
	( 65.51, 71.99) --
	( 65.76, 72.43) --
	( 66.01, 72.87) --
	( 66.26, 73.32) --
	( 66.51, 73.78) --
	( 66.76, 74.23) --
	( 67.01, 74.70) --
	( 67.26, 75.17) --
	( 67.51, 75.64) --
	( 67.76, 76.12) --
	( 68.01, 76.61) --
	( 68.27, 77.11) --
	( 68.52, 77.62) --
	( 68.77, 78.13) --
	( 69.02, 78.65) --
	( 69.27, 79.19) --
	( 69.52, 79.73) --
	( 69.77, 80.29) --
	( 70.02, 80.86) --
	( 70.27, 81.44) --
	( 70.52, 82.04) --
	( 70.77, 82.65) --
	( 71.02, 83.27) --
	( 71.27, 83.91) --
	( 71.52, 84.57) --
	( 71.77, 85.25) --
	( 72.02, 85.94) --
	( 72.27, 86.65) --
	( 72.52, 87.38) --
	( 72.77, 88.13) --
	( 73.02, 88.90) --
	( 73.27, 89.69) --
	( 73.52, 90.51) --
	( 73.77, 91.34) --
	( 74.02, 92.20) --
	( 74.27, 93.09) --
	( 74.52, 94.00) --
	( 74.77, 94.93) --
	( 75.02, 95.89) --
	( 75.27, 96.87) --
	( 75.52, 97.87) --
	( 75.77, 98.91) --
	( 76.02, 99.97) --
	( 76.27,101.05) --
	( 76.52,102.15) --
	( 76.77,103.27) --
	( 77.02,104.41) --
	( 77.27,105.58) --
	( 77.52,106.76) --
	( 77.77,107.96) --
	( 78.02,109.18) --
	( 78.27,110.40) --
	( 78.52,111.64) --
	( 78.77,112.89) --
	( 79.02,114.14) --
	( 79.27,115.40) --
	( 79.53,116.67) --
	( 79.78,117.93) --
	( 80.03,119.20) --
	( 80.28,120.46) --
	( 80.53,121.72) --
	( 80.78,122.98) --
	( 81.03,124.22) --
	( 81.28,125.46) --
	( 81.53,126.69) --
	( 81.78,127.91) --
	( 82.03,129.12) --
	( 82.28,130.32) --
	( 82.53,131.50) --
	( 82.78,132.67) --
	( 83.03,133.83) --
	( 83.28,134.98) --
	( 83.53,136.11) --
	( 83.78,137.23) --
	( 84.03,138.33) --
	( 84.28,139.43) --
	( 84.53,140.51) --
	( 84.78,141.58) --
	( 85.03,142.63) --
	( 85.28,143.68) --
	( 85.53,144.71) --
	( 85.78,145.74) --
	( 86.03,146.75) --
	( 86.28,147.74) --
	( 86.53,148.73) --
	( 86.78,149.70) --
	( 87.03,150.65) --
	( 87.28,151.60) --
	( 87.53,152.52) --
	( 87.78,153.43) --
	( 88.03,154.31) --
	( 88.28,155.18) --
	( 88.53,156.03) --
	( 88.78,156.86) --
	( 89.03,157.66) --
	( 89.28,158.43) --
	( 89.53,159.18) --
	( 89.78,159.91) --
	( 90.03,160.61) --
	( 90.28,161.28) --
	( 90.53,161.92) --
	( 90.79,162.53) --
	( 91.04,163.12) --
	( 91.29,163.67) --
	( 91.54,164.19) --
	( 91.79,164.69) --
	( 92.04,165.14) --
	( 92.29,165.56) --
	( 92.54,165.95) --
	( 92.79,166.31) --
	( 93.04,166.63) --
	( 93.29,166.92) --
	( 93.54,167.16) --
	( 93.79,167.37) --
	( 94.04,167.53) --
	( 94.29,167.66) --
	( 94.54,167.75) --
	( 94.79,167.79) --
	( 95.04,167.79) --
	( 95.29,167.74) --
	( 95.54,167.65) --
	( 95.79,167.51) --
	( 96.04,167.33) --
	( 96.29,167.11) --
	( 96.54,166.83) --
	( 96.79,166.51) --
	( 97.04,166.15) --
	( 97.29,165.75) --
	( 97.54,165.31) --
	( 97.79,164.82) --
	( 98.04,164.29) --
	( 98.29,163.72) --
	( 98.54,163.12) --
	( 98.79,162.48) --
	( 99.04,161.81) --
	( 99.29,161.10) --
	( 99.54,160.35) --
	( 99.79,159.58) --
	(100.04,158.77) --
	(100.29,157.93) --
	(100.54,157.06) --
	(100.79,156.16) --
	(101.04,155.23) --
	(101.29,154.26) --
	(101.54,153.26) --
	(101.79,152.22) --
	(102.05,151.15) --
	(102.30,150.05) --
	(102.55,148.90) --
	(102.80,147.71) --
	(103.05,146.49) --
	(103.30,145.22) --
	(103.55,143.92) --
	(103.80,142.58) --
	(104.05,141.19) --
	(104.30,139.76) --
	(104.55,138.31) --
	(104.80,136.81) --
	(105.05,135.29) --
	(105.30,133.73) --
	(105.55,132.15) --
	(105.80,130.55) --
	(106.05,128.93) --
	(106.30,127.31) --
	(106.55,125.68) --
	(106.80,124.05) --
	(107.05,122.44) --
	(107.30,120.84) --
	(107.55,119.26) --
	(107.80,117.70) --
	(108.05,116.18) --
	(108.30,114.71) --
	(108.55,113.28) --
	(108.80,111.90) --
	(109.05,110.58) --
	(109.30,109.32) --
	(109.55,108.11) --
	(109.80,106.98) --
	(110.05,105.92) --
	(110.30,104.92) --
	(110.55,103.99) --
	(110.80,103.13) --
	(111.05,102.33) --
	(111.30,101.60) --
	(111.55,100.94) --
	(111.80,100.33) --
	(112.05, 99.78) --
	(112.30, 99.28) --
	(112.55, 98.83) --
	(112.80, 98.42) --
	(113.05, 98.06) --
	(113.31, 97.73) --
	(113.56, 97.43) --
	(113.81, 97.16) --
	(114.06, 96.90) --
	(114.31, 96.67) --
	(114.56, 96.46) --
	(114.81, 96.25) --
	(115.06, 96.04) --
	(115.31, 95.84) --
	(115.56, 95.64) --
	(115.81, 95.44) --
	(116.06, 95.23) --
	(116.31, 95.01) --
	(116.56, 94.79) --
	(116.81, 94.55) --
	(117.06, 94.30) --
	(117.31, 94.04) --
	(117.56, 93.76) --
	(117.81, 93.46) --
	(118.06, 93.16) --
	(118.31, 92.83) --
	(118.56, 92.49) --
	(118.81, 92.13) --
	(119.06, 91.76) --
	(119.31, 91.37) --
	(119.56, 90.97) --
	(119.81, 90.56) --
	(120.06, 90.13) --
	(120.31, 89.69) --
	(120.56, 89.24) --
	(120.81, 88.78) --
	(121.06, 88.32) --
	(121.31, 87.85) --
	(121.56, 87.37) --
	(121.81, 86.89) --
	(122.06, 86.40) --
	(122.31, 85.91) --
	(122.56, 85.42) --
	(122.81, 84.93) --
	(123.06, 84.44) --
	(123.31, 83.95) --
	(123.56, 83.47) --
	(123.81, 82.98) --
	(124.06, 82.50) --
	(124.31, 82.02) --
	(124.57, 81.55) --
	(124.82, 81.08) --
	(125.07, 80.62) --
	(125.32, 80.16) --
	(125.57, 79.71) --
	(125.82, 79.26) --
	(126.07, 78.82) --
	(126.32, 78.39) --
	(126.57, 77.96) --
	(126.82, 77.53) --
	(127.07, 77.11) --
	(127.32, 76.70) --
	(127.57, 76.29) --
	(127.82, 75.89) --
	(128.07, 75.49) --
	(128.32, 75.09) --
	(128.57, 74.70) --
	(128.82, 74.31) --
	(129.07, 73.92) --
	(129.32, 73.53) --
	(129.57, 73.15) --
	(129.82, 72.77) --
	(130.07, 72.39) --
	(130.32, 72.01) --
	(130.57, 71.63) --
	(130.82, 71.26) --
	(131.07, 70.88) --
	(131.32, 70.51) --
	(131.57, 70.14) --
	(131.82, 69.77) --
	(132.07, 69.40) --
	(132.32, 69.04) --
	(132.57, 68.68) --
	(132.82, 68.32) --
	(133.07, 67.97) --
	(133.32, 67.62) --
	(133.57, 67.28) --
	(133.82, 66.94) --
	(134.07, 66.61) --
	(134.32, 66.29) --
	(134.57, 65.97) --
	(134.82, 65.66) --
	(135.07, 65.36) --
	(135.32, 65.07) --
	(135.58, 64.78) --
	(135.83, 64.50) --
	(136.08, 64.22) --
	(136.33, 63.96) --
	(136.58, 63.70) --
	(136.83, 63.46) --
	(137.08, 63.21) --
	(137.33, 62.98) --
	(137.58, 62.75) --
	(137.83, 62.53) --
	(138.08, 62.32) --
	(138.33, 62.11) --
	(138.58, 61.91) --
	(138.83, 61.71) --
	(139.08, 61.52) --
	(139.33, 61.33) --
	(139.58, 61.15) --
	(139.83, 60.98) --
	(140.08, 60.80) --
	(140.33, 60.63) --
	(140.58, 60.47) --
	(140.83, 60.31) --
	(141.08, 60.15) --
	(141.33, 59.99) --
	(141.58, 59.84) --
	(141.83, 59.69) --
	(142.08, 59.54) --
	(142.33, 59.40) --
	(142.58, 59.25) --
	(142.83, 59.11) --
	(143.08, 58.98) --
	(143.33, 58.84) --
	(143.58, 58.71) --
	(143.83, 58.58) --
	(144.08, 58.45) --
	(144.33, 58.33) --
	(144.58, 58.21) --
	(144.83, 58.09) --
	(145.08, 57.98) --
	(145.33, 57.87) --
	(145.58, 57.76) --
	(145.83, 57.66) --
	(146.08, 57.56) --
	(146.33, 57.47) --
	(146.58, 57.38) --
	(146.84, 57.30) --
	(147.09, 57.22) --
	(147.34, 57.14) --
	(147.59, 57.07) --
	(147.84, 57.00) --
	(148.09, 56.94) --
	(148.34, 56.87) --
	(148.59, 56.82) --
	(148.84, 56.76) --
	(149.09, 56.71) --
	(149.34, 56.66) --
	(149.59, 56.62) --
	(149.84, 56.57) --
	(150.09, 56.53) --
	(150.34, 56.49) --
	(150.59, 56.45) --
	(150.84, 56.41) --
	(151.09, 56.37) --
	(151.34, 56.34) --
	(151.59, 56.30) --
	(151.84, 56.27) --
	(152.09, 56.23) --
	(152.34, 56.20) --
	(152.59, 56.16) --
	(152.84, 56.13) --
	(153.09, 56.10) --
	(153.34, 56.07) --
	(153.59, 56.03) --
	(153.84, 56.00) --
	(154.09, 55.97) --
	(154.34, 55.94) --
	(154.59, 55.91) --
	(154.84, 55.88) --
	(155.09, 55.85) --
	(155.34, 55.82) --
	(155.59, 55.80) --
	(155.84, 55.77) --
	(156.09, 55.74) --
	(156.34, 55.72) --
	(156.59, 55.69) --
	(156.84, 55.67) --
	(157.09, 55.65) --
	(157.34, 55.63) --
	(157.59, 55.61) --
	(157.84, 55.59) --
	(158.10, 55.57) --
	(158.35, 55.55) --
	(158.60, 55.53) --
	(158.85, 55.51) --
	(159.10, 55.49) --
	(159.35, 55.48) --
	(159.60, 55.46) --
	(159.85, 55.45) --
	(160.10, 55.43) --
	(160.35, 55.42) --
	(160.60, 55.41) --
	(160.85, 55.39) --
	(161.10, 55.38) --
	(161.35, 55.37) --
	(161.60, 55.36) --
	(161.85, 55.35) --
	(162.10, 55.34) --
	(162.35, 55.33) --
	(162.60, 55.32) --
	(162.85, 55.31) --
	(163.10, 55.30) --
	(163.35, 55.29) --
	(163.60, 55.29) --
	(163.85, 55.28) --
	(164.10, 55.27) --
	(164.35, 55.27) --
	(164.60, 55.26) --
	(164.85, 55.26) --
	(165.10, 55.25) --
	(165.35, 55.25) --
	(165.60, 55.25) --
	(165.85, 55.24) --
	(166.10, 55.24) --
	(166.35, 55.24) --
	(166.60, 55.24) --
	(166.85, 55.24) --
	(167.10, 55.23) --
	(167.35, 55.23) --
	(167.60, 55.23) --
	(167.85, 55.23) --
	(168.10, 55.23) --
	(168.35, 55.23) --
	(168.60, 55.23) --
	(168.85, 55.23) --
	(169.10, 55.23) --
	(169.36, 55.23) --
	(169.61, 55.23) --
	(169.86, 55.23) --
	(170.11, 55.23) --
	(170.36, 55.23) --
	(170.61, 55.23);
\end{scope}
\begin{scope}

\definecolor[named]{drawColor}{rgb}{0.00,0.00,0.00}

\draw[color=drawColor,line cap=round,line join=round,] ( 52.47, 50.40) -- (164.21, 50.40);

\draw[color=drawColor,line cap=round,line join=round,] ( 52.47, 50.40) -- ( 52.47, 44.40);

\draw[color=drawColor,line cap=round,line join=round,] ( 80.40, 50.40) -- ( 80.40, 44.40);

\draw[color=drawColor,line cap=round,line join=round,] (108.34, 50.40) -- (108.34, 44.40);

\draw[color=drawColor,line cap=round,line join=round,] (136.27, 50.40) -- (136.27, 44.40);

\draw[color=drawColor,line cap=round,line join=round,] (164.21, 50.40) -- (164.21, 44.40);

\node[color=drawColor,anchor=base,inner sep=0pt, outer sep=0pt, scale=  1.00] at ( 52.47, 26.40) {0.2};

\node[color=drawColor,anchor=base,inner sep=0pt, outer sep=0pt, scale=  1.00] at ( 80.40, 26.40) {0.1};

\node[color=drawColor,anchor=base,inner sep=0pt, outer sep=0pt, scale=  1.00] at (108.34, 26.40) {0};

\node[color=drawColor,anchor=base,inner sep=0pt, outer sep=0pt, scale=  1.00] at (136.27, 26.40) {0.1};

\node[color=drawColor,anchor=base,inner sep=0pt, outer sep=0pt, scale=  1.00] at (164.21, 26.40) {0.2};

\draw[color=drawColor,line cap=round,line join=round,] ( 48.00, 55.23) -- ( 48.00,175.31);

\draw[color=drawColor,line cap=round,line join=round,] ( 48.00, 55.23) -- ( 42.00, 55.23);

\draw[color=drawColor,line cap=round,line join=round,] ( 48.00, 85.25) -- ( 42.00, 85.25);

\draw[color=drawColor,line cap=round,line join=round,] ( 48.00,115.27) -- ( 42.00,115.27);

\draw[color=drawColor,line cap=round,line join=round,] ( 48.00,145.29) -- ( 42.00,145.29);

\draw[color=drawColor,line cap=round,line join=round,] ( 48.00,175.31) -- ( 42.00,175.31);

\node[rotate= 90.00,color=drawColor,anchor=base,inner sep=0pt, outer sep=0pt, scale=  1.00] at ( 36.00, 55.23) {0};

\node[rotate= 90.00,color=drawColor,anchor=base,inner sep=0pt, outer sep=0pt, scale=  1.00] at ( 36.00, 85.25) {2};

\node[rotate= 90.00,color=drawColor,anchor=base,inner sep=0pt, outer sep=0pt, scale=  1.00] at ( 36.00,115.27) {4};

\node[rotate= 90.00,color=drawColor,anchor=base,inner sep=0pt, outer sep=0pt, scale=  1.00] at ( 36.00,145.29) {6};

\node[rotate= 90.00,color=drawColor,anchor=base,inner sep=0pt, outer sep=0pt, scale=  1.00] at ( 36.00,175.31) {8};
\end{scope}
\begin{scope}

\definecolor[named]{drawColor}{rgb}{0.00,0.00,0.00}

\draw[color=drawColor,line cap=round,line join=round,] (108.34, 50.40) -- (108.34,180.67);
\end{scope}

%% file: phylogenetic_appendix1.tex
\section{Algorithms for computing medians and means} \label{sec:mmalg}
The approximation algorithms for computing medians and means were introduced by~\cite{bacak} and we refer the interested reader therein for the proofs of convergence and further details. The algorithms rely upon a well-known optimization technique called the proximal point method. Interestingly, the algorithm for computing the mean can be alternatively justified via the law of large numbers due to \cite{sturm}, as was independently observed by \cite{bacak} and \cite{miller}.

\subsection{Algorithms for computing medians}
Let us first describe the algorithm for computing a median of a given set $t_1,\dots,t_K\in\ts_n.$

We set $x_0\as t_1$ and suppose that at the $i$-th iteration we have an approximation $x_i \in \ts_n$ of $\Psi\l(\ol{t}\r).$ To find $x_{i+1},$ a tree $t_k$ is selected from our set of trees $t_1,\dots,t_K$ at random and we define $x_{i+1}$ as a point on the geodesic between $x_i$ and $t_k$. (In other words $x_{i+1}$ is a convex combination of $x_i$ and $t_k.$) The position of $x_{i+1}$ on this geodesic is determined by a parameter $t_i\in[0,1],$ which is computed at each iteration. By this procedure, we obtain a sequence of trees $x_1,x_2,\dots$ which is known converge to a median of $t_1,\dots,t_K.$

\begin{alg}[Computing median, random order version] \label{alg:medianran}
Let $x_0\as t_1.$ At each step $i\in\nat_0,$ choose randomly $r_i\in\{1,\dots,K\}$ according to the uniform distribution and put
\begin{equation} \label{eq:medran}
 x_{i+1}\as \l(1-\eta_i\r)x_i + \eta_i t_{r_i},
\end{equation}
with $\eta_i$ defined by
\begin{equation*} 
\eta_i\as\min\left\{1,\frac{1}{(i+1)d\left(t_{r_i},x_i\right)}\right\},
\end{equation*}
for each $i\in\nat_0.$ 
\end{alg}
It is important to insist on the uniform distribution on the set $\{1,\dots,K\},$ that is, no tree of $t_1,\dots,t_K$ be privileged. Only then we obtain a sequence of trees $x_1,x_2,\dots$ which converges to a median of $t_1,\dots,t_K.$

\subsection{Algorithms for computing means}
Computing the mean is similar to the computation of the median. As a matter of fact it only differs in the coefficients determining the position of $x_{i+1}$ on the geodesic from $x_i$ to $t_k.$  Again, let $t_1,\dots,t_K\in\ts_n$ be a finite set of trees from $\ts_n.$ The following approximation algorithms generate a sequence of trees $x_1,x_2,\dots$ from $\ts_n$ which converges to $\Xi\l(\ol{t}\r).$  At each iteration a tree $t_k$ is selected at random and we obtain the following algorithm.
\begin{alg}[Computing mean, random order version] \label{alg:meanran}
Let $x_0\as t_1$ and at each step $i\in\nat_0,$ choose randomly $r_i\in\{1,\dots,K\}$ according to the uniform distribution and put
\begin{equation*}
 x_{i+1}\as \frac{1}{i+1}x_i + \frac{i}{i+1} t_{r_i}.
\end{equation*}
\end{alg}

The above algorithms have also their deterministic counterparts, where we choose the trees from the input set in a cyclic order instead of randomly; see~\cite{bacak}. Even though both random and cyclic versions converge to the same value, there is no theorem on which one converges faster. Our computational studies however suggest that the random versions are better.